\documentclass[a4paper, 10pt]{article}

\usepackage{amsmath} % AMS Math Package
\usepackage{amssymb}	% Math symbols such as \mathbb
\usepackage{graphicx} % Allows for eps images
\usepackage{multicol} % Allows for multiple columns
\usepackage{color}
\usepackage{caption}
\usepackage{subcaption}
\usepackage{cite}
\usepackage{array}
\usepackage{multirow}
\usepackage[font={small}]{caption}
\usepackage{booktabs}
\usepackage{emptypage}
\usepackage{tikz}
\usepackage{authblk}
\usepackage{etoolbox}
\usepackage{lmodern}
\usepackage{setspace}
\usepackage[T1]{fontenc}
\usepackage[utf8]{inputenc}
\usepackage[english]{babel}
\usepackage{float}
\usepackage{geometry}
\usepackage{color}
\usepackage{hyperref}
\DeclareMathOperator*{\argmaxB}{argmax} 

\makeatletter % Need for anything that contains an @ command 

\let\baraccent=\= % rename builtin command \= to \baraccent
\renewcommand{\=}[1]{\stackrel{#1}{=}} % for putting numbers above =
\newcommand{\defeq}{\mathrel{\mathop:}=} %for define equal symbol :=

\hypersetup{
    colorlinks,
    citecolor=black,
    filecolor=black,
    linkcolor=black,
    urlcolor=black
}

\usepackage{subcaption}
\begin{document}

\title{\bf Statistical Physics and  Representations\\ in Real and Artificial Neural Networks}

\author{S. Cocco$^1$, R. Monasson$^2$, L. Posani$^1$, S. Rosay$^3$,   J. Tubiana$^2$ \\
$^1$ Laboratoire de Physique Statistique, Ecole Normale Sup\'erieure and CNRS, PSL Research, Sorbonne Universit\'es UPMC, 24 rue Lhomond,75005 Paris, France\\
$^2$ Laboratoire de Physique Th\'eorique, Ecole Normale Sup\'erieure and CNRS, PSL Research, Sorbonne Universit\'es UPMC, 24 rue Lhomond,75005 Paris, France\\
$^3$ Cognitive Neuroscience, SISSA, via Bonomea 265, Trieste, Italy}
\maketitle

\begin{abstract}
This document presents the material of two lectures on statistical physics and neural representations, delivered by one of us (R.M.) at the Fundamental Problems in Statistical Physics XIV summer school in July 2017. In a first part, we consider the neural representations of space (maps) in the hippocampus. We introduce an extension of the Hopfield model, able to store multiple spatial maps as continuous, finite-dimensional attractors. The phase diagram and dynamical properties of the model are analyzed. We then show how spatial representations can be dynamically decoded using an effective Ising model capturing the correlation structure in the neural data, and compare applications to data obtained from hippocampal multi-electrode recordings and by (sub)sampling our attractor model. In a second part, we focus on the problem of learning data representations in machine learning, in particular with artificial neural networks. We start by introducing data representations through some illustrations. We then analyze two important algorithms, Principal Component Analysis and Restricted Boltzmann Machines, with tools from statistical physics.
\end{abstract}

\tableofcontents

%\newpage

\section{Introduction}

In the early 80's, statistical physicists proved that ideas issued from their field could lead to substantial advances in other disciplines. Simulated Annealing, a versatile optimization procedure in which a fictitious sampling temperature is decreased until the minimum (ground state) of a cost function is reached, had major impact in applied computer science and engineering \cite{Kirkpatrick83}. Attractor neural network models for memories \cite{Hopfield82}, soon analytically solved with spin-glass techniques \cite{AGS87}, emerged as one major conceptual tool in computational neuroscience. From a theoretical point of view, it became rapidly clear that statistical physics offered a powerful framework to deal with problems outside physics, in particular in computer science and theoretical neuroscience, involving many random, heterogeneous, strongly interacting components, which had remained very hard to tackle so far.

The purpose of the present document is to present some applications of statistical physics ideas and tools to the understanding of high-dimensional representations in neural networks. How the brain represents and processes information coming from the outside world is a central issue of computational neuroscience \cite{Rieke97}. Experimental progress in electrophysicological and optical recordings make now possible to record the activity of populations of tens to thousands of neural cells in behaving animals, opening the way to study this question with unprecedented access to data and to ask new questions about brain operation on large scales \cite{Sompolinsky14}. Concomittantly, machine learning algorithms, largely based on artificial neural network architectures, have recently achieved spectacular performance in a variety of fields, such as image processing, or speech recognition/production \cite{LeCun2015}. How these machines produce efficient representations of the data and of their underlying distributions is a crucial question \cite{Bengio2013}, far from being understood \cite{Coveney16}. Profound similarities seem to emerge between the representations encountered in real and artificial neural networks \cite{poggio16} and between the questions raised in both contexts \cite{Ganguli12}.

It is utterly hard to cover recent advances in such a diverse and vivid field, and the task is impossible in two lectures of two hours each. The material gathered here merely reflects the interests and, presumably, the ignorance of the authors more than anything else. The present notes focus on two applications of statistical physics to the study of neural representations in the contexts of computational neuroscience and machine learning. The first part is motivated by the representation of spaces, i.e. multiple environments, in hippocampal place-cell networks.  An extension of Hopfield's attractor neural network to the case of finite-dimensional attractors is introduced and its phase diagram and dynamical properties, such as diffusion within one attractor or transitions between distinct attractors, are analyzed. We also show that effective, functional Ising models fitted from hippocampal  multi-electrode recordings (limited to date to few tens of neurons) or from 'neural' data generated by spatially subsampling our model, share common features with our abstract model, and can be used to decode and  to track the evolution of spatial representations over time. In a second part, we move to representations of data by machine learning algorithms. Special emphasis is put on two aspects: low-dimensional representations achieved by principal component analysis, and compositional representations, produced by restricted Bolztmann machines combining multiple features inferred from data. In both cases, we show how statistical physics helps unveil the different properties of these representations, and the role of essential control parameters.

\section{Representation of space(s) in the hippocampus: model}
\subsection{Context and background}

\subsubsection{Zero-dimensional attractors: Hopfield model of associative memory}

Statistical Mechanics and Neuroscience are not so far apart as they may seem at first sight. Indeed, brains are made of billions of neurons that are connected together. In many cases, brain functions are thought to be the outcome of collective states. This makes it a good playground for Statistical Mechanics. Here, we will focus on one particular brain function: memory. 

In 1949, D. Hebb had the visionary intuition that  memory could correspond to the retrieval of certain activity patterns in a network of interconnected neurons  \cite{Hebb49}. This \emph{attractor hypothesis} goes as follows: (1) what is memorized are attractors of the network, \emph{i.e.} activity states stable under the dynamical evolution rule; hence, recalling a memory corresponds to retrieving its activity pattern; (2) attractors are stored in the network couplings $J_{ij}$ that govern the network dynamics and stable states; (3) a possible way to make an arbitrary pattern an attractor is to 'wire together neurons that fire together' in that pattern (the so-called 'Hebb rule'). 

In 1982, J.J. Hopfield \cite{Hopfield82} proposed a model based on Hebb's ideas in the case of zero-dimensional, or, equivalently, point attractors. This model, known as the Hopfield model,  is strongly inspired by statistical physics models used in the context of the magnetic systems, such as the Ising model. It consists of a number $N$ of binary neurons ${\{s_i\}_{i=1\dots N}=\pm 1}$ and stores a number $P$ of configurations ${\{\xi_i^\mu\}_{i=1\dots N,\mu=1\dots p}=\pm1 }$ (point attractors), e.g. independently and uniformly drawn at random. The synaptic couplings that allow these configurations to be attractors are given by the Hebb rule:
\begin{equation}
\label{eq:hebb}
J_{ij}=\frac1N\sum\limits_{\mu=1}^P\xi_i^\mu\xi_j^\mu\ \ \forall i,j\ .
\end{equation}
The last thing to define is the dynamics of the network. In the original paper \cite{Hopfield82}, time was discretized and, at each time step $t$, neurons responded deterministically to their local fields, through the updating rule: 
\begin{equation}\label{dyna}
s_i^{t+1} = \text{sign} \big( \sum_j J_{ij} s_j ^t \big) \ .
\end{equation} 
Later studies, e.g. \cite{AGS85}, incorporated the possibility of stochasticity in the response, through a noise parameter $T$, so that the system obeyed detailed balance for the Gibbs distribution associated to the Hamiltonian
\begin{equation} \label{ej}
E_J[{\bf s}]=-\sum\limits_{i<j}J_{ij}s_i s_j \ 
\end{equation}
at 'temperature' $T$.

In terms of biological relevance, the Hopfield model is of course extremely schematic. Yet, it captures many fundamental and robust aspects of neurons (in particular their linear summation of inputs, combined with a thresholding effect) and network (synaptic coefficients with values affected by the activity through the Hebb rule), while remaining, to a large extent, analytically tractable. The Hopfield model aroused a great excitement in the Statistical Mechanics community during the 80's, since it shared many common points with frustrated and disordered magnetic systems. Tools and methods from the statistical physics of disordered systems that had just been developed in the field of spin glasses \cite{mezard87} could therefore  be used to derive analytically the properties of the Hopfield model \cite{AGS85}. The first question was  to check whether the patterns ${\{\xi_i^\mu\}_{i=1\dots N,\mu=1\dots p}}$ were indeed attractive fixed points of the dynamics in eqn (\ref{dyna}). The answer turned out to be positive (up to a small fraction of the $N$ neurons) for small enough values of $T$ and of the ratio $\alpha=P/N$ (in the double limit $N,P\to\infty$), i.e. for not too strong noise and memory load.  Many aspects of this model were studied and refined, in particular to make it more biologically realistic. The reader is kindly referred to \cite{Amit89} for a detailed presentation of the literature. Rather, we will focus in an extension of this model to a different kind of attractors, that is finite-dimensional attractors.

\subsubsection{Place cells in the rodent hippocampus}

\begin{figure}
  \begin{minipage}[c]{0.5\textwidth}
    \includegraphics[width=\textwidth]{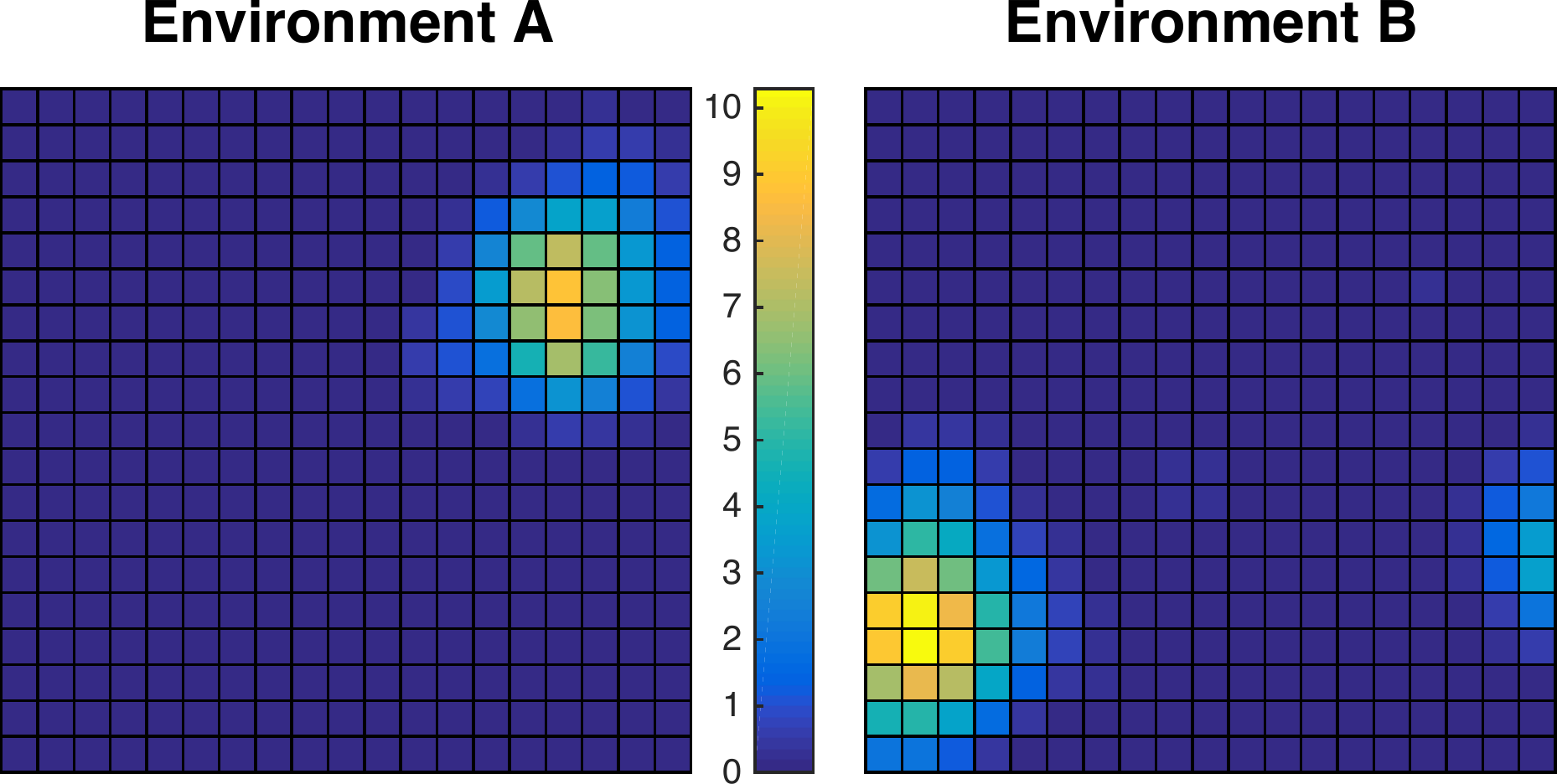}
  \end{minipage}\hfill
  \begin{minipage}[c]{0.45\textwidth}
    \caption{
       Remapping of place field for one recorded place cell in CA1 for a living rat exploring two square environments A and B with identical shapes. The size of the environments is $60\times 60$ cm. The figure reports the  average firing rate of the recorded cell when the rat is in each of the $3\times 3$-cm spatial bins; values in Hz, see color bar. Data from experiment by K. Jezek et al \cite{Jezek11}.      } 
    \label{fig:remappingca1}
  \end{minipage}
\end{figure}

Let us now turn to real brains, more specifically, how space is represented in the brain \cite{Moser08review}. Experimentalists use small electrodes that, implanted in the brain of awake animals, are able to record the simultaneous activity of a population of \emph{single neurons}. 
%This way, Hebb's attractor idea has received support from many neural systems. 
In particular, in a brain area called hippocampus, O'Keefe \& Dostrovsky have discovered the existence of 'place cells' when recording in rodents freely moving in an enclosure \cite{OKeefeDostrovsky71}. These neurons have the surprising property that they fire only when the animal is physically located in a precise region of space, hence their name. The region of activity corresponding to a place cell in the environment defines its 'place field'. In the CA3 area of the hippocampus, a region with strong recurrent connection between pyramidal cells, the different place fields attached to a given place cell across different environments visited by the rodent seem to be totally uncorrelated --- a property called global 'remapping'. In another hippocampal area, called CA1, remapping of place fields from one environment to another is generally weaker; the change in the activity of a place cell is characterized mainly by a modulation of its firing rate, a phenomenon called rate remapping, though global changes of the place fields as in CA3 may also be observed for some cells, see Fig.~\ref{fig:remappingca1}.

For many reasons, the hippocampus --- more precisely its subregion CA3 --- is often supposed to work as a \emph{continuous} attractor neural network \cite{TrevesRolls94,Tsodyks99}. It means that the attractors are not point configurations (of zero dimension, as in the Hopfield model), but attractors in one \footnote{The one-dimensional case would correspond to linear corridors.} or two dimensions\footnote{Though there is experimental evidence that place cells code also for rich contextual information \cite{SmithMizumori06}, we consider only the spatial correlate of place-cell activity in the present document.} : each manifold corresponds to an environment, \emph{i.e.} the collection of activity configurations of the hippocampal neural population characterizing the set of all positions in that environment. Apart from the dimensionality of the attractors, place cells share common points with the Hopfield model, such as the absence of correlation between attractors due to random remapping, and the Hebb rule that has some biological counterparts. Hence, it is appealing to extend the Hopfield model to continuous attractors.

%In addition to place cells, other neurons have an activity related to physical space and are thought to be attractor systems: the head-direction cells, the grid cells... See \cite{Knierim12} for a review.

\subsection{A model for memorizing D-dimensional attractors (spatial maps)}
\label{secmodel}

We thus introduce a model for place cells in one- or two-dimensional spaces (the extension to higher dimensions is straightforward). As an extension of the Hopfield model, our model is based on binary neurons; other models with real-valued neural variable, e.g. firing rates, can be found in literature \cite{BattagliaTreves98,Samsonovich97}. 
%Contrary to the Hopfield model original's notation, we assume that the neural activity variables take value $s_i=0,1$ instead of $\pm1$.
The $N$ place cells are modeled by binary units $s_i$ equal to 0 (silent state) or 1 (active state)\footnote{We will hereafter use indifferently the terms "neuron", "place cell" and "spin", from the analogy with magnetic systems.}. These neurons interact together through excitatory couplings $J_{ij}$. Moreover, they interact with inhibitory interneurons, whose effect is to maintain the total activity of the place cells to a fraction $f$ of active cells (global inhibition). We also assume that there is some stochasticity in the response of the neurons, controlled by a noise parameter $T$. All these assumptions come down to considering that the network states are distributed according to the Gibbs distribution associated to the Hamiltonian (\ref{ej}), restricted to configuration of spins $\bf s$ such that
\begin{equation}\label{constraint1}
\sum_i s_i=f N \ .
\end{equation}
%\begin{equation}
% Z_J(T)=\sum\limits_{\underset{\sum\limits_i s_i=fN}{\boldsymbol s \text{ such that }}}\exp(-E_J(\boldsymbol s)/T)\ ,
%\end{equation}
% where the ``energy'' (in the thermodynamic sense) of a configuration $\boldsymbol s$ reads
%\begin{equation}
% E_J(\boldsymbol s)=-\sum\limits_{i<j}J_{ij} s_i s_j\ .
%\end{equation}

\begin{figure}
  \begin{minipage}[c]{0.4\textwidth}
    \includegraphics[width=\textwidth]{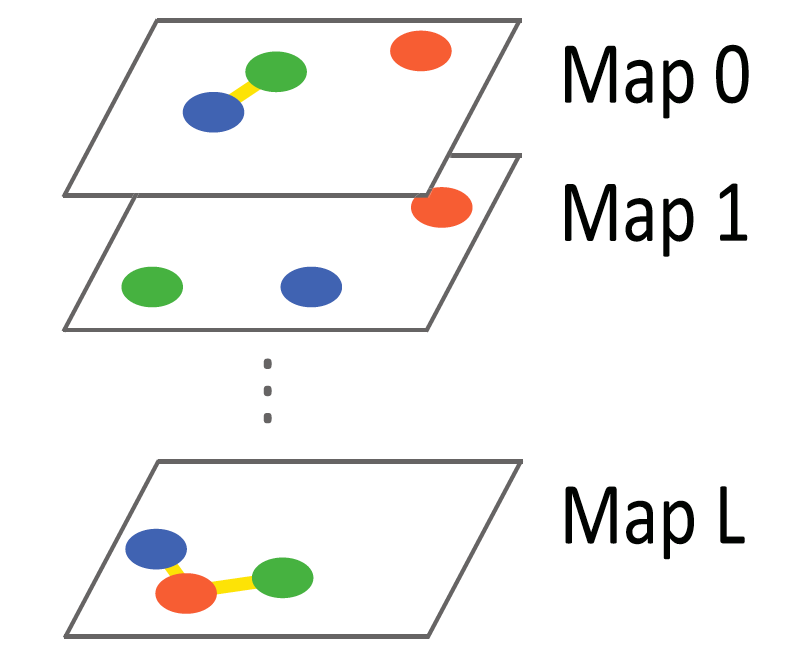}
  \end{minipage}\hfill
  \begin{minipage}[c]{0.5\textwidth}
    \caption{
       Remapping and connectivity rule in the model, illustrated with three units and ${L+1}$ two-dimensional environments. The place field centers of the units are displayed respectively in red, blue and green. Thick yellow lines indicate the excitatory couplings between cells with neighbouring place fields in each environment. These place fields overlap; here, for the sake of clarity, only the centers of the place fields are represented. 
    } 
    \label{fig:remappingmodel}
  \end{minipage}
\end{figure}

We want to store $L+1$ environments in the coupling matrix. We call place field a position of space where a place cell preferentially fires. An environment  $\ell$ is defined as a random permutation $\pi^\ell$ of the $N$ neurons' place fields (assuming that the place fields are regularly arranged on a grid). This models the experimentally observed remapping of place fields from one map to the other\footnote{In this basic version of the model, every place cells have place fields in every environments. The possibility of silent cells has been taken into account  \cite{Monasson13}}. With this definition, an environment is said to be stored when activity patterns localized in this environment are stable states of the dynamics. In other words, the configurations where active neurons have neighbouring place fields in this environment are equilibrium states. To make this possible, we assume a Hebbian prescription for the couplings $J_{ij}$ that is a straightforward extension of the Hopfield synaptic matrix to the case of quasi-continuous attractors. This rule is illustrated in Figure~\ref{fig:remappingmodel}, and is mathematically described as follows:
\begin{itemize}
 \item additivity: $J_{ij}=\sum\limits_{\ell=0}^LJ_{ij}^\ell$ where the sum runs over all the environments.
\item potentiation of excitatory couplings between units that may become active together when the animal explores the environment:
\begin{equation}\label{rule1}
J^\ell_{ij} = 
\frac 1N \ \hbox{\rm if} \ d^\ell_{ij} \le d_c\ , \quad 
0 \ \hbox{\rm if} \ d^\ell_{ij} > d_c  \ ,
\end{equation}
where $d^\ell_{ij}$ is the distance between the place-field centers of $i$ and $j$ in the environment $\ell$; for instance, in dimension $D=1$, $d^\ell_{ij}=\frac 1N|\pi^\ell(i)-\pi^\ell(j)|$. $d_c$ represents the distance over which place fields overlap. In practice, it is chosen so that, in each environment, each neural cell is coupled to a fraction $w$ of the other cells (its neighbours); in dimension $D=1$ again, we may choose $d_c=\frac w2$. The $\frac 1N$ factor in eqn (\ref{rule1}) ensures that the total input received by a cell remains finite as $N$ goes to infinity, a limit case in which exact calculations become possible \cite{lebowitz66}.
\end{itemize}

\subsection{Replica theory and phase diagram}
\label{sec:phasediag}

The aim of this calculation is to study the stable states of the network, and to find under which conditions these stable states correspond to a set of active neurons whose corresponding place fields are nearby in  one of the environments.  In other words, we want to know for which parameter values the Hebbian synaptic matrix (\ref{rule1}) ensures the retrieval of the stored maps. The system under study enjoys both disordered (due to the random allocation of place fields in each map) and frustrated (from the competition between excitatory synapses and the global inhibition) interactions. 

We start by computing the free energy of the system,
\begin{equation}
 F=-T\log Z_J(T)\ , \quad \text{where}\quad Z_J(T)=\sum_{\boldsymbol s\ \text{with constraint}\ (\ref{constraint1})}\exp(-E_J(\boldsymbol s)/T)\ .
\end{equation}
This quantity depends a priori on the realization of the random permutations in each map. We assume that, in the large $N$ limit, the free energy is self-averaging: its particular value for a given realization of the disorder is typically close to its average over all possible realizations of the disorder, which is thus a good approximation of $F$. The randomness of the remapping process is thus a key hypothesis for the model to be tractable. To compute the average of the logarithm of $Z_J(T)$ we use the replica method \cite{mezard87}: we first compute the $n^{th}$ moment of $Z_J(T)$, and then compute its first derivative with respect to $n\to 0$. 

Since we are interested in configurations where the place fields of the active neurons are spatially concentrated in one of the environments, we arbitrarily select one of the environments (called ``reference environment'') and do the averaging over the remaining $L$ other permutations; details about the calculation can be found in \cite{Monasson13}. This choice is totally arbitrary because the difference between environments is eventually averaged out. In the reference environment, neurons are indexed in the same order as their place fields, which allows us to move from a microscopic activity configuration $\boldsymbol s$ to a macroscopic activity density over continuous space
\begin{equation}\label{density_ave}
\rho(x) \equiv\lim _{\epsilon \to 0}\lim _{N \to \infty} \; \frac 1{\epsilon N} \sum_{(x-\frac \epsilon 2)N\le i < (x+\frac \epsilon 2)N}\overline{ \langle s _i \rangle_J }\ ,
\end{equation}
where the overbar denotes the average over the random remappings while the brackets correspond to the average over the fast noise. For simplicity, we have assumed that the environment is one-dimensional here, but the above formula can easily be extended to higher dimensions. Note that our model is  analytically tractable  in the large $N$ limit (thermodynamic limit), as each unit has an infinite number of neighbours it weakly interacts with. The mean-field approximation  therefore becomes exact in the sense that order parameters such as the density in  (\ref{density_ave}) exhibits no fluctuation when $N\to\infty$; however, contrary to standard mean field approaches, these order parameters depend on space \cite{lebowitz66}. 

\begin{figure}
\centering
    \includegraphics[width=.9\textwidth]{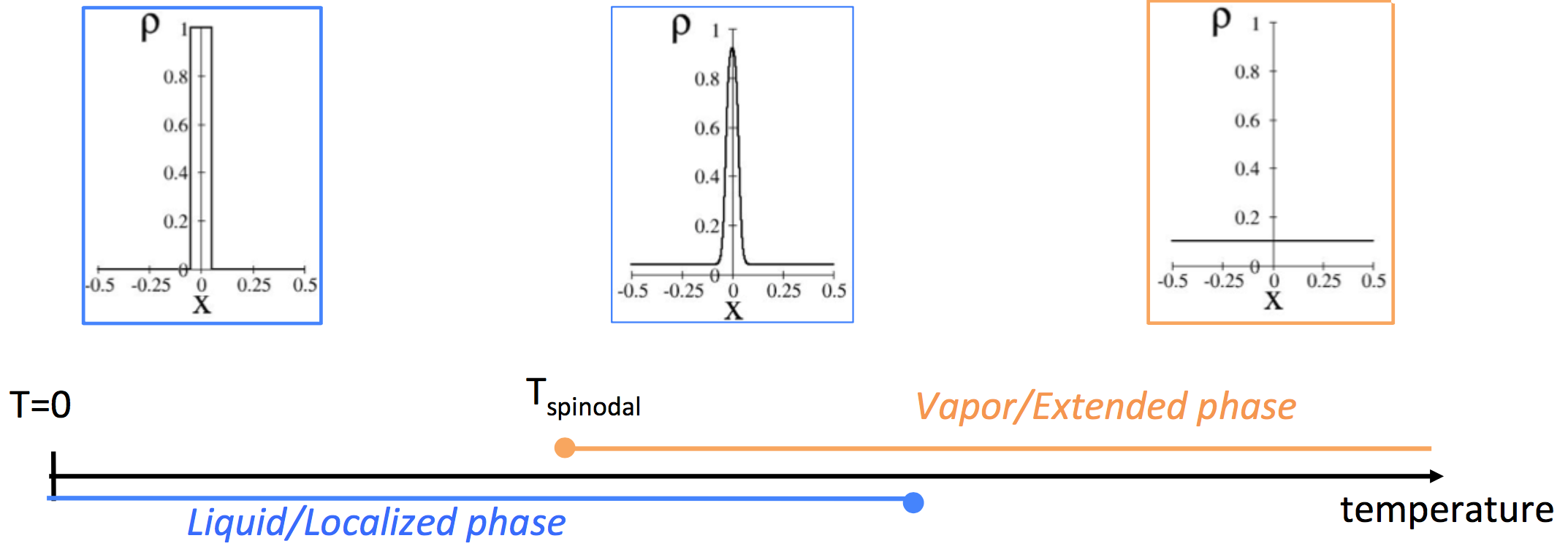}
    \caption{Phase diagram of Lebowitz \& Penrose's theory of the lquid/vapor transition, in dimension $D=1$ (periodic boundar conditions). Insets show the density of particles $\rho(x)$ as a function of position over space, $x\in[0;1]$. Parameters: $f=0.1$, $w=0.05$. Note the coexistence between the homogeneous and bump states at intermediate temperatures. The location of the bump is arbitrary.
           } 
    \label{fig:lp}
\end{figure}

The case of a single (reference) environment, i.e. $L=0$, is strongly reminiscent of the theory of the liquid-vapor transition by Lebowitz and Penrose \cite{lebowitz66}: the continuous
translational symmetry is spontaneously broken at low enough temperature, i.e. $\rho(x) \ne f$, and a liquid drop (bump of high density fluid) is surrounded by low-density vapor, see Fig.~\ref{fig:lp}. this bump can then freely diffuse, and describes a finite-dimensional continuum of ground states.

In the $L>0$ case, the $n^{th}$ moment of $Z_J(T)$ is given by equation (20) in \cite{Monasson13}. The averaged term depends on the configurations $\boldsymbol s^1,\cdots\boldsymbol s^n$ only through the overlap matrix with entries $q^{ab}=\frac 1N  \boldsymbol s^a \cdot\boldsymbol s^b$ \cite{mezard87,Monasson13}. 
Then, to perform the $n\to 0$ limit, we consider the replica symmetric Ansatz, which assumes that the overlaps $q^{ab}$ take a single value (over all replica indices $a\ne b$),
\begin{equation}\label{defqab}
q \equiv \frac 1N \sum_{j} \overline{ \langle s _i \rangle_J ^2}\ ,
\end{equation}
which is the Edwards-Anderson parameter of spin glasses, characterizing the site-to-site fluctuations of the spin magnetizations \cite{EdwardsAnderson75}. This Ansatz is generally valid at high-enough temperature, when the Gibbs measure defined on the energy landscape is not too rough, see below. It allows us to compute the free energy as a function of the order parameters $\rho(x)$, $\mu(x)$ (chemical potential conjugated to $\rho(x)$), $q$ and $r$ (conjugated to $q$). $\mu(x)$ and $r$ have simple interpretations. The effective field acting on neuron $i$, whose place field is located in $x=i/N$ in the reference environment, is the sum of two terms: a 'signal' contribution $\mu(x)$ coming from neighboring neurons $j$ in the retrieved map (through the couplings $J^0_{ij}$), and a Gaussian noise, of zero mean and variance $\alpha\, r$ coming from the other maps $\ell\ge 1$ (see Fig. 14 in \cite{Monasson13}). Here, $\alpha\equiv L/N$ denotes the load of the memory.

The four order parameters (two scalar, two functional) fulfilll the following saddle-point equations obtained through extremization of the free-energy functional,
\begin{eqnarray}\label{eq:saddlepoint1}
r&=&2(q-f^2)\sum\limits_{k\geq 1}\left[\frac{k \pi}{\sin(k\pi w)}-\beta(f-q)\right]^{-2}\ , \quad 
q=\int\mathrm{d}x\int\mathrm{D}z\; \big[1+e^{-\beta z\sqrt{\alpha r}-\beta\mu(x)}\big]^{-2}\ ,\nonumber\\
\rho (x)&=&\int\mathrm{D}z\; \big[1+e^{-\beta z\sqrt{\alpha r}-\beta\mu(x)}\big]^{-1}\ , \quad 
\mu(x)=\int\mathrm{d}y\, J_w(x-y)\, \rho(y)+\lambda\ ,
\end{eqnarray}
where $\beta\equiv1/T$,  $Dz=\exp(-z^2/2)/\sqrt{2\pi}$ is the Gaussian measure, and $\lambda$ is determined to enforce the fixed activity level constraint ${\int\mathrm{d}x\,\rho(x)=f}$, see (\ref{constraint1}). The precise expression of $r$ depends on the eigenvalue spectrum of the $J^0_{ij}$ matrix. Changing the hard cut-off $d_c$ to a smooth e.g. exponential decay of the coupling with the distance $d_{ij}^\ell$ between the place-field centers would change the expression for $r$, but would not affect the overall behaviour of the model.

\begin{figure}
  \begin{minipage}[c]{0.6\textwidth}
    \includegraphics[width=\textwidth]{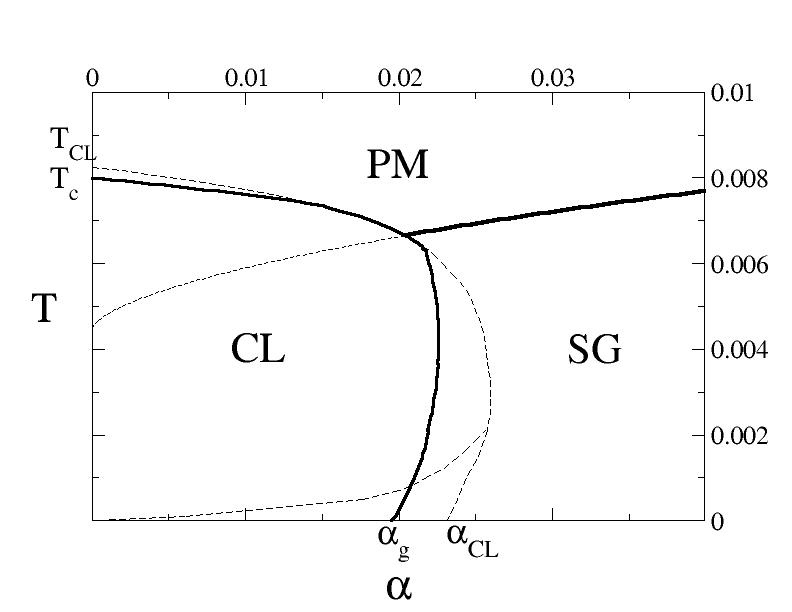}
  \end{minipage}\hfill
  \begin{minipage}[c]{0.33\textwidth}
    \caption{
    Phase diagram in the $(\alpha,T)$ plane in $D=1$ with $f=0.1$ and $w=0.05$. Thick lines: transition between phases. Dashed-dotted line: ${T_\text{PM}(\alpha)}$. Thin dashed line: CL phase's longitudinal stability regions. Dotted line: CL phase's RSB line. ${\alpha_\text{CL}}$: storage capacity at ${T=0}$ of the replica-symmetric clump phase. ${\alpha_g}$: CL-SG transition load at ${T=0}$.  ${T_\text{CL}}$: temperature of loss of stability of the clump  at ${\alpha=0}$. ${T_c}$: CL-PM transition temperature at ${\alpha=0}$. ${T_\text{PM}=T_\text{PM}(\alpha=0)}$ (see text). 
           } 
    \label{fig:phasediag}
  \end{minipage}
\end{figure}

We find three distinct solutions to these coupled equations:
\begin{itemize}
\item a paramagnetic phase (PM), corresponding to high levels of noise $T$, in which the average local activity is uniform over space, $\rho(x)=f$, and neurons are essentially uncorrelated, $q=f^2$.  
\item a 'clump' phase (CL), where the activity depends on space, {\em i.e.} $\rho(x)$ varies with $x$ and is localized in the reference environment. This phase corresponds to the 'retrieval phase' where the environment is actually memorized. In fact, all the $L+1$ environments are memorized since any of them could be chosen as the reference environment. Note that the value of $x$ (center of the bump of activity) is totally arbitrary, as all positions are equivalent after averaging over the permutations.
\item  a glassy phase (SG), corresponding to large loads $\alpha$, in which the local activity $\langle s_i\rangle$ varies from neuron to neuron ($q>f^2$), but does not cluster around any specific location in space in any of the environments ($\rho(x)=f$ after averaging over remappings). In this SG phase the crosstalk between environments is so large that none of them is actually stored in the network activity.
In the SG phase, contrary to the CL phase, no environment is memorized. This is the 'black-out catastrophe' \cite{Amit89} already described in the Hopfield model, in which retrieval also takes place in an all-or-nothing fashion.
\end{itemize}
 We now need to determine which solution is selected as functions of $\alpha$, $T$, that is, the phase of lowest free energy that will be thermodynamically favored, as well as the domains of existence (stability) of those three phases against longitudinal and replicon modes \cite{deAlmeidaThouless78}, and the transition lines between them. To study the stability, we write the Hessian of the free energy and study its eigenvalues in the longitudinal and replicon sectors. Then, the transition between two phases is the line where the free energies in both phases equalize. We have done these calculations in the one-dimensional case, as detailed in ref. \cite{Monasson13}. The outcome is the phase diagram shown in Fig.~\ref{fig:phasediag}, displaying the three phases domains in the $(\alpha,T)$ plane:

\begin{itemize}
 \item the paramagnetic solution exists for all $\alpha$, $T$ and is stable for ${T>T_\text{PM}(\alpha)}$ displayed with the dot-dashed line in Fig.~\ref{fig:phasediag}.
\item the glassy phase exists for ${T<T_\text{PM}(\alpha)}$, and is always replica-symmetry broken; we expect replica symmetry breaking to be continuous in this region, as in the celebrated Sherrington-Kirkpatrick model  \cite{mezard87}.
\item the longitudinal stability of the clump phase is computed numerically and shown with the thin dashed line in Fig.~\ref{fig:phasediag}. The clump is stable against replicon modes except in a little low-$T$ high-$\alpha$ region (dotted line). An interesting feature of the clump phase stability domain is the reentrance of the high-$\alpha$ boundary. 
\end{itemize}
We have checked this analytically-derived phase diagram by Monte Carlo simulations. For a detailed comparison of this phase diagram with the one of the Hopfield model, and how the dimensionality of the attractors plays a role, see  \cite{theseSophie}.

\subsection{Dynamics within one map and transitions between maps}

The phase diagram above informs us on the stable states of the model.For moderate temperature and memory load, i.e. in the CL phase, thermodynamically stable states have an activity spatially localized somewhere in one of the maps. But this does not constrain \emph{which} map is retrieved and \emph{where} in this map, i.e. what is the position intersecting the place fields of the active cells. Indeed, under the influence of noise, the bump of activity can move around in a given map, and also jump to another map. We have studied both these dynamics within one map and between maps, respectively in \cite{Monasson14} and \cite{Monasson15}.

\begin{figure}
\hspace*{\fill}%
  \subcaptionbox{Session A\label{reference-a}}{\includegraphics[width=1.5in]{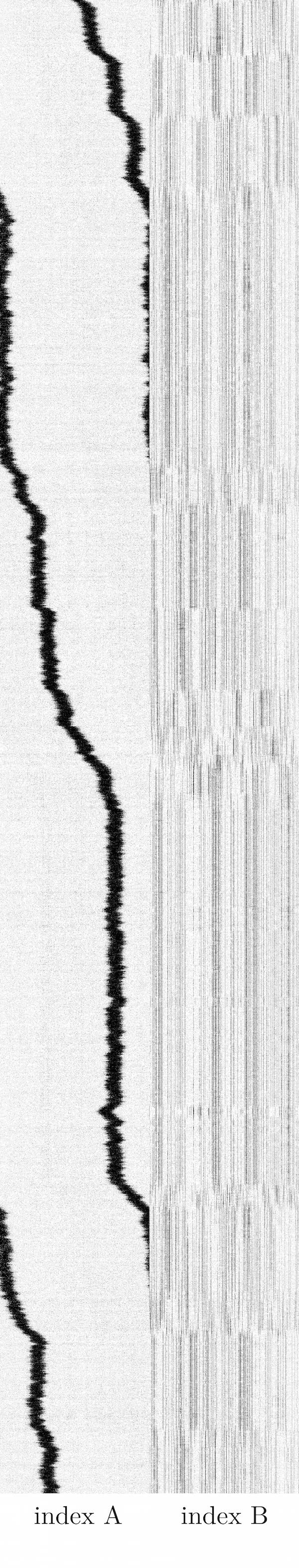}}\hfill%
  \subcaptionbox{Session B\label{reference-b}}{\includegraphics[width=1.5in]{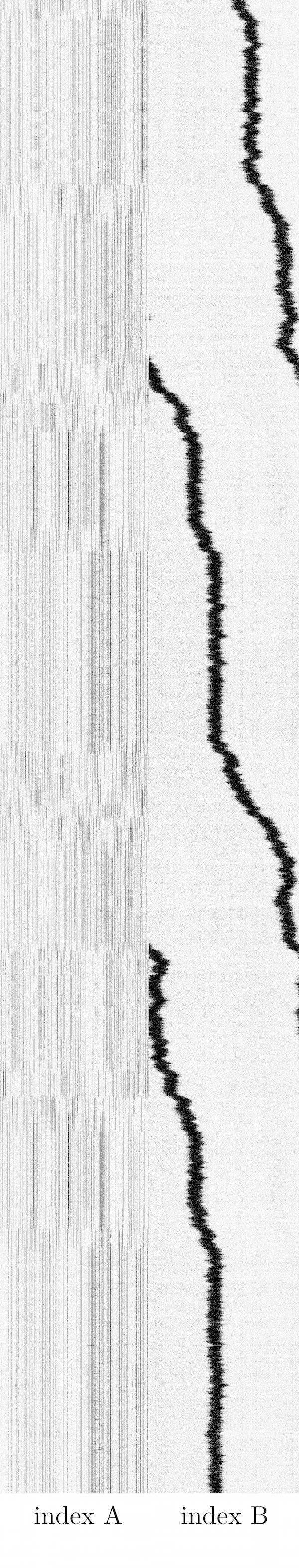}}\hfill%
  \subcaptionbox{Test Session\label{test}}{\includegraphics[width=1.5in]{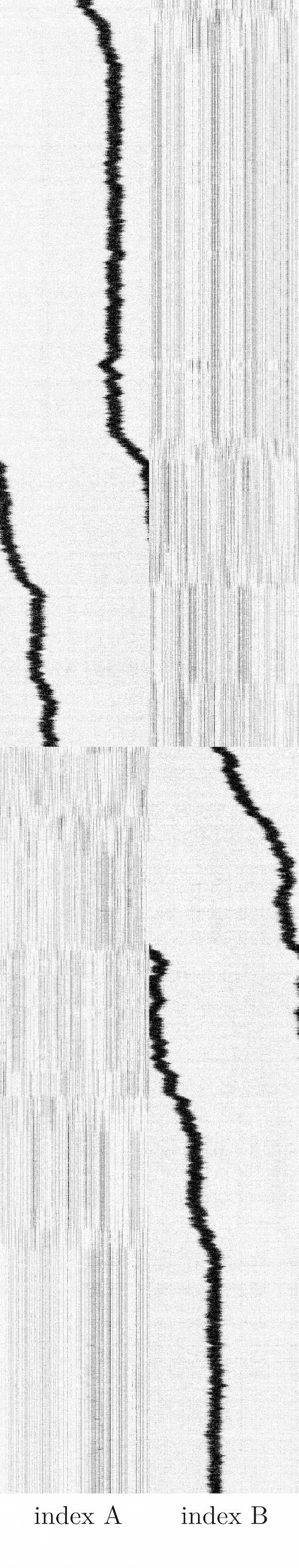}}%
  \hspace*{\fill}%
  \caption{Monte Carlo simulation sessions of our memory model in the case of two 1D environments (random permutations), denoted by A and B.  X-axis: states of the system ${\bf s}(t)$ (black dots correspond to active neurons $s_i=1$ and white dots to silent cells, $s_i=0$), with neurons ordered in increasing order of their place field centers in the A (left part of columns) or B (right part of columns) permutations. Y-axis: time in MC rounds, increasing from top to down. The bump is forced to move rightwards with an external force, see \cite{Monasson14}. 
   In columns {\bf (a)} and {\bf (b)}, the system is initialized with a localized bump of activity in environments, respectively, A and B.
  Column {\bf (c):} Test simulations composed of the second halves of  simulations reported in {\bf (a)} and {\bf (b)} used for decoding purposes, see text. Parameter values: $T=0.006$, $N=1000$, $w = 0.05$, $f= 0.1$.}
\label{montecarlo}
\end{figure}

Within one map, we have shown formally that, in the case of a single continuous attractor (one map, \emph{i.e.} $\alpha=0$), the bump of activity behaves like a quasi-particle with little deformation. This quasi-particle undergoes a pure diffusion with a diffusion coefficient that can be computed exactly from first principle, i.e. from the knowledge of microscopic flipping rates of spins in Monte Carlo simulations. The diffusion coefficient scales as $1/N$, see Eq.~(31) in \cite{Monasson14}. When imposing a force (spins) on the spins, see Section \ref{secforce}, the activity changes so as to move the bump. An illustration is shown in Fig.~\ref{montecarlo}. It can be shown analytically that the mobility of the bump and its diffusion coefficient obey the Stokes-Einstein relation. 

In the presence of multiple maps, the disorder due to the presence of multiple maps stored in the couplings creates an effective free-energy landscape for the bump of activity in the reference environment. The free-energy barriers scale typically as $\sqrt N$, and are correlated over space length of the order of the bump size, see \cite{Monasson14}. In one dimension, the bump therefore effectively undergoes Brownian motion in the Sinai potential, with strongly activated diffusion. In higher dimension, diffusion is facilitated with respect to the 1D case, as can be observed in simulations.

In addition to moving in the reference environment, the bump can also spontaneously jump between maps. Fast transitions between maps, evoked by light inputs, have been observed by K. Jezek and colleagues in the so-called 'teleportation experiment' \cite{Jezek11}. Understanding these transitions in the framework of a simple model provides insight on the mechanisms involved in the biological system. Map-to-map transitions can be studied with replica theory again, but in a more subtle framework, where solutions with non-uniform activities in two maps (and not only one as in eqn. (\ref{eq:saddlepoint1}) above) are searched for. There are two scenarios for spontaneous transitions between spatial representations, see Fig. 4 in \cite{Monasson15}: 
\begin{itemize}
\item through a mixed state, which gives bumps of activity in both maps; these bumps are weaker than the one in the CL phase in a single reference environment. This scenario is preferred (has lower free-energy cost) at low $T$. Transitions take place at special 'confusing' positions in both environments, where both maps locally resemble most.
\item through a non-localized state, i.e. through the PM phase. Owing to the liquid-vapor analogy, the bump of activity in map A evaporates, and then condensates in map B.  This scenario is preferred
(has lower free-energy cost) at high $T$ (but sufficiently low to make the CL phase thermodynamically favorable with respect to PM, see phase diagram in Fig.~\ref{fig:phasediag}).
\end{itemize}

We show in Fig.~\ref{figbarriere}A the rate of transitions between maps computed from Monte Carlo simulations, see Supplemental Material in \cite{Monasson15} for details. We observe that the rate increases with temperature, and diminishes with the load and the system size. According to Langer's nucleation theory \cite{Langer69}, we expect the rate to be related to the free-energy barrier $\Delta F$ between CL phase in environment A and the CL phase in environment B through (see formula 3.37 in \cite{Langer69}),
\begin{equation}\label{lang6}
R = \kappa \; \sqrt{\frac{T}{2\pi\,N\,|\lambda_-|}}\; {\cal V}\; \exp\big( - N\, \Delta F/T\big) \ ,
\end{equation}
where $\kappa$ is the growth rate of the unstable mode at the transition state, $\lambda_-$ is the unique negative eigenvalue of the Hessian of the free-energy at the transition state, ${\cal V}$ is the volume of the saddle-point subspace (resulting from the integral over continuous degrees of freedom leaving the saddle point configuration globally
unchanged). Hence, we expect the measures of the rates obtained for different system size to collapse onto each other upon the following rescaling:
\begin{equation}\label{sca}
R \to  - \frac{\log\big( R\, \sqrt N \big) }N \ .
\end{equation}
This scaling is nicely confirmed by numerics, see Fig.~\ref{figbarriere}B. We observe the collapse to a limiting curve, related to the barrier height through $\Delta F/T$. Note that $\Delta F$ is itself a function of temperature $T$, calculated in \cite{Monasson15}.

\begin{figure}
\begin{center}
\includegraphics[width=0.8\linewidth]{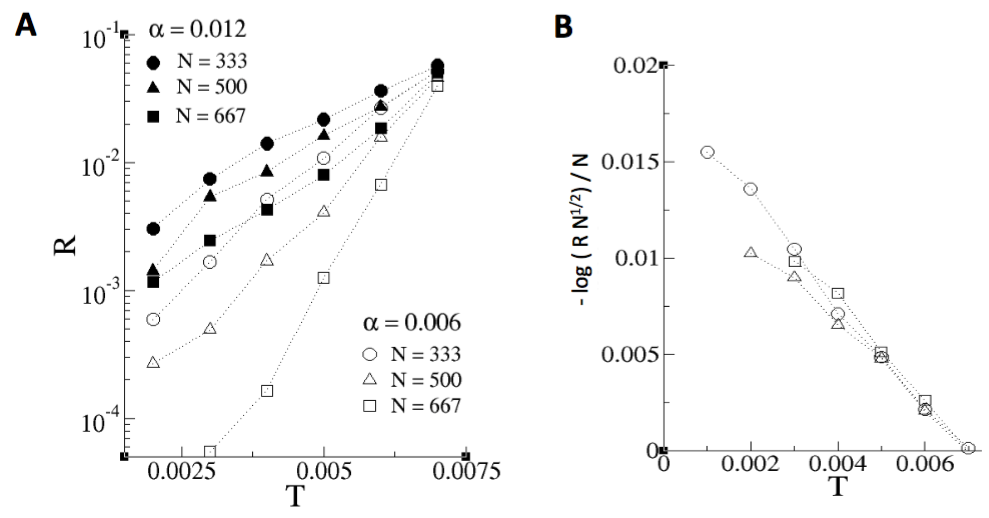}  
\end{center}
\caption{Map-to-map spontaneous transitions. {\bf A.} rate $R$ of transitions computed from Monte Carlo simulations, as a function of the temperature $T$, see \cite{Monasson15}. Parameters: $f = 0.1$, $w = 0.05$. {\bf B.} Replotting the same data as in panel {\bf A} for $\alpha= 0.006$ after transformation in eqn (\ref{sca}) allows us to estimate the ratio of the free-energy barrier over temperature, see eqn (\ref{lang6}).}
\label{figbarriere}
\end{figure}

\section{Representation of space(s) in the hippocampus: decoding and data analysis}

\subsection{Decoding neural representations with effective Ising networks}
The problem of decoding which brain state is internally represented from the observation of neural activity has a natural application in experiments involving the simultaneous recording of a neural population \cite{cocco2017functional}. The \textit{decoding} problem can be tackled by learning statistical properties of a known set of brain states and classifying new observations accordingly, a problem which is deeply connected to high-dimensional classification in machine learning. An example is provided by the environmental memory in the hippocampus. Activity is recorded while the animal explores a set of environments. Each environment is associated to a memorized cognitive map, and these \textit{reference sessions} are then used to learn models of the activity associated to each map. In turn, these models can be used to decode activity recorded during a controlled \textit{test session}, i.e. decide which map this activity is associated to. During the test session, environmental cues are manipulated by the experimentalist, and decoding the internal response to the change of experimental conditions allows us to investigate the induced dynamics over internal representations \cite{Jezek11, posani2017functional}.

We can tackle the decoding problem by inferring a probability density function over the neural patterns ${\bf s}$ for each brain state $M$, $P({\bf s} | M)$\footnote{For definiteness,  we will hereafter consider the neural configuration at a given time, ${\bf s} = \{s_1, s_2, \ldots , s_N\}$, to be the set of binarized neural activities of the neurons $i$=1...N under consideration, with $s_i=0$ if neuron $i$ is silent, $1$ if it is active, see Section \ref{secising} for more details.}. These probability distributions  can be used to decode the internal state $M$ given an observation (a neural pattern) ${\bf s}$ in the test session. More precisely, $M$ is decoded by maximizing the log-likelihood
\begin{equation}
\mathcal{L}(M |{\bf s}) = \log P({\bf s} |M)\ .
\end{equation}
This inference framework relies on the definition of a parametric probability function, whose parameters are inferred by solving the corresponding \textit{inverse problem} from reference data. According to the max-entropy principle, our choice is to use the family of graphical models \cite{jaynes1957information, wainwright2008graphical, MacKayBook} as parametric probabilistic functions. Depending on the reference sample size and/or the complexity of representations we can invert the Independent model, which accounts for the diferent average activations of neurons in different brain states, or make a step further and include correlations between neural activity, defining an Ising model for each state $M$.
\begin{equation}\label{pisi}
P({\bf s} |M) =  \frac{\exp{\left(\sum_{i} h^{M}_{i} s_{i} + \sum_{i<j} J_{ij}^{M} s_{i}s_{j} \right)}}{\mathcal{Z}^{M}(h,J)}
\end{equation}
where ${\cal Z}^M$ is a normalization constant
\begin{equation}\label{z}
{\cal Z}^M(h,J)= \sum_{\bf s} \exp{\left(\sum_{i} h^{M}_{i} s_{i} + \sum_{i<j} J_{ij}^{M} s_{i}s_{j} \right)}\ .
\end {equation}

The core steps of the Ising decoding procedure are:

\begin{enumerate}
\item {\em Reference session.} For each brain state $M$, (a) collect samples of neural pattern  in a known brain state $M$ (reference session), and compute the frequencies  $p_{i}^{M}$ and pairwise joint frequencies $p_{ij}^{M}$ of the recorded neurons; (b) find the Ising model that reproduces the same quantities on average, i.e.  such that $\left< s_{i} \right> = p_{i}^{M}$ and $\left< s_{i}s_{j} \right> = p_{ij}^{M}$, where $\left< \cdot \right>$ denotes the average over the probability distribution $P({\bf s} |M)$. This is a highly non-trivial computational problem, reviewed in Section \ref{secising}.
\item{\em Test session.}  Given a neural pattern from the test session ${\bf s}^{t}$, compute the log-likelihood of each brain state, and decode the internal state as the most likely one
\begin{equation}
M^{t} = \argmaxB_{M} \ \mathcal{L}(M |{\bf s}^{t})\ .
\end{equation}
\end{enumerate}
Within this framework we can therefore decode the neural representation from the observed neural pattern. This procedure has been applied to experimental data from the hippocampus, showing good performance in retrieving the explored environment from neural activity \cite{posani2017functional}. Similar procedures have been successfully applied to other brain regions, see for instance \cite{tavoni2017functional,Tavoni2016b, schneidman2006weak, Stevenson08}. 

Before applying the decoding procedure in the context of the representations of space, let us review how the effective Ising model in eqn (\ref{pisi}) can be fitted from data.

\subsection{ Inference of effective Ising  models  from data }
\label{secising}
We discuss in this Section the problem of  the inference of a graphical model from  data \cite{ackley1985learning,schneidman2006weak,wainwright2008graphical,aurell2012inverse,cocco2011adaptive}.
Data are defined here as a  set of $B$ recorded configurations of $N$   variables,  ${\bf s}^b = \{s_1^b, s_2^b, \ldots , s_N^b\}$, with $i=1,\ldots ,N$;  $s_i^b$ denotes the value of the variable at site $i$ in configuration $b$.  Variables $s_i^b$  can be real valued, binary, or multi-categorical (Potts state). Two cases of great practical interest are:
\begin{itemize}
\item Neurons can be described by binary variables $s_i=0,1$, expressing whether they are silent or active, i.e. emit a spike.  Spiking times obtained from multi-electrode recordings  \cite{mcnaughton83b,meister1994,schneidman2006weak,Peyrache09} can be processed into a series of $B=T/\Delta t$ recorded neural configurations ${\bf s}^b$, by dividing the recording time $T$ in small time windows $b=1...B$ of duration $\Delta t$; then, $s_i^b$ is equal to 1 if neuron emits one or more spikes in time bin $b$, to 0 if it remains silent. 
\item Amino acid $s_i$ at site $i$ in a protein sequence can take 20 different values. Data configurations are sequences from a familiy of homologous proteins, assumed to share a common 3D fold and biological function, collected in protein databases \cite{durbin1998biological,finn2016pfam,de2013emerging,morcos2011direct,balakrishnan2011learning}.
\end{itemize}

For the sake of simplicity, we assume hereafter that variables take binary values, $s=0,1$. We further assume that the distribution over configurations $\bf s$ of  $N$ such variables is given by the Ising model defined in eqn (\ref{pisi}); to lighten notations, we will drop the $M$ subscript hereafter. The model is parametrized by $ N$  fields $h_i$ and $\frac 12 N( N -1)$  couplings $J_{ij}$.

%\section{Bayesian framework}
We assume that the different data configurations are independently  drawn from $P({\bf s}|h,J)$ in eqn (\ref{pisi}). Hence, the probability of the data reads
\begin{equation}
\prod _{b=1}^B P\big({\bf s}^b |h,J \big) = \exp \left[- B \ S\big(h ,J \big)\right]
\end{equation}
where the  cross-entropy $S$ is
\begin{equation}
\label{entropy}
S(h,J )= \log Z(h,J) - \sum _{i} h_i\;  p_i - \sum _{i<j}  J_{ij}\; p_{ij} 
 \end{equation}
 depends on the data through the single-site and pairwise  frequencies
 \begin{equation}
p_i =\frac 1B \sum_b s_i^b \quad \text{and}\quad p_{ij} =\frac 1B \sum_b s_i^b \, s_j^b  \ .
\end{equation}
The best values for the fields and the couplings are the one minimizing $S$. While $S$ is a convex function of its arguments, the minimum is not guaranteed to be finite. For instance, the minimum of $S$ is realized at $J_{12}=-\infty$ when  neurons  1 and 2 never spike together ($p_{12}=0$). This problem can
be avoided by including a prior, also called regularization, over the fields and couplings. Usual regularization schemes include adding the $L_1$ and/or $L_2$ norms of the couplings, to avoid small nonzero couplings or couplings with very large, unrealistic values. Another regularization scheme consists of imposing a small rank for the coupling matrix, see Section \ref{sechopfieldinv}.

The computational problem in the minimization of eqn (\ref{entropy}) is the calculation of the partition function $ Z(h,J) $, which is generally intractable as  it involves a summation over the $2^N$ configuration of the systems. Some methods to solve the inverse Ising problem bypass the calculation of $Z$, such has   Boltzmann Machine algorithm \cite{ackley1985learning}, the Pseudo-Likelihood approximations \cite{wainwright2008graphical,aurell2012inverse},  the minimum probability flow \cite{sohl2011new}, or resort to approximate expressions for $Z$, e.g. mean field \cite{Opper2001}, high-temperature expansions \cite{sessak2009small}, and  adaptive cluster  expansions \cite{cocco2011adaptive,cocco2012}.
     
Once the Ising model has been inferred, it can be used for various tasks: 
\begin{itemize} 
\item Extract structural information on the connectivity/coupling matrix between the variables. In the case of neurons, this {\em functional connectivity} is not physiological (synaptic), but is an effective set of couplings depending on the brain state \cite{friston2011,tavoni2017functional,cocco2017functional}. In protein covariation analysis, it has been shown that  large couplings often coincide with amino acids in contact on the three-dimensional structure of the protein \cite{morcos2011direct,de2013emerging}.
\item Use the inferred model to score new configurations and decide if they are compatible with the data in the training data set. We  will see a direct application in Sections~\ref{secmod2} and \ref{secdata2}. 
\item Generate new configurations  through Monte Carlo simulations. This can be very useful to obtain in silico data with the same features as the ones in the training set, for instance, new proteins with the same structure or function as natural proteins.
\end{itemize}

\subsection{Back to model: the subsampling problem} \label{secmod2}

The theoretical model for spatial memory in place-cell populations from Section \ref{secmodel} shows remarkable features compatible with the recall of brain states at the level of population activity \cite{monasson2013crosstalk, monasson2014crosstalk}. The existence of a \textit{clump phase}, in which the system is maintained in a local minimum of self-sustained localized activity, is compatible with the attractor-neural-network (ANN) general paradigm of cognitive functions being represented by collective states of the neural network. The presence of spontaneous transitions from one representation to another is consistent with the flickering phenomena triggered by weak inputs observed in the rat hippocampus \cite{Jezek11}, where spatial memory is thought to be stored and retrieved \cite{tsodyks1995associative, o1971hippocampus, o1978hippocampal}. However, theoretical results were obtained in the limit of very large systems, which is also the case of real brain regions ($\sim 10^{9}$ neurons), while electrophysiological setups permit to record simultaneously a much smaller ($<10^{2}$) number of neurons. It is natural to wonder to what extent such a small number of neurons could provide information about the collective state of the whole population.

Hereafter, we describe an attempt to draw a parallel between experimental conditions of multi-array recordings and the theoretical model for environment memory in the attractor neural network framework. We first design a Monte Carlo simulation that mimics an experiment with two memorized environments, referred to as A and B. We simulate single-environment {\em reference sessions} by forcing the activity to explore local minima corresponding to the memorized environments in a system with a relatively large number ($N=1,000$) of neurons. We then address the question if a small, randomly selected, set of neurons (here, $N_{sam}=33$ over 1,000) could provide enough information to perform the decoding procedure and infer the time course of the spatial representations from neural activity. As place cells are non topographical, i.e. cells that are physically nearby in the hippocampus can have distant place fields, recording a spatially located population of cells can be thought of to be equivalent to a random subsample in the place-field abstract space.

The decoding task is finally performed on the test session using Ising and independent models learned from the reference sessions, and their decoding capability is tested in a classification problem on a test session composed by samples from both states. The relationship between true and inferred coupling is then analyzed.

\subsubsection{Simulations: constructing reference and test sessions}
\label{secforce}
Monte Carlo simulations are conducted as follows:
\begin{itemize}

\item First we define two 1D environments, hereby referred to as A and B, through their two random place-field permutations, denoted by $\pi^{A}$, $\pi^B$. 

\item From these two environments, two coupling matrices $J^{M}$, $M \in \left\{ A, B \right\}$, are created using learning prescription described in eqn (\ref{rule1}):
\begin{equation}
\label{neuron}
J^{M}_{ij} \defeq \left\{ 
\begin{array}{l l l}
    \frac 1N & \quad \text{if } & \frac1N \left | \pi^{M}(i) -  \pi^{M}(j) \right | \leq \frac{w}{2} \ , \\
    0 & \ & \text{otherwise}\ .
  \end{array} \right.
\end{equation}
\item A unique coupling matrix $J$ is then constructed as point-sum of the two single-environment matrices: $J_{ij}=J^{A}_{ij} + J^{B}_{ij}$.
\item simulations are performed, with $n = 10^{4}$ Monte Carlo steps, each one starting from an initial neuronal condition  localized in one of the two reference environments $M$. To maintain the total activity  constant, we select, at each algorithm step, one active spin $s_{i} = 1$ and one silent spin $s_{j} = 0$. The flip trial is then defined as the joint flip of these spins.
\item an additional small force is added to make the bump exhaustively explore the one dimensional map, by an asymmetric term in the energy. This results in a left-right asymmetry in the Monte Carlo acceptance rule:
\begin{equation}
\Delta E = \sum_{k \neq i,j} (J_{ik} - J_{jk}) s_{k} + A^{M}(i,j)
\end{equation}
with $A^{M}(i,j)$ being a right-pulling force in the environment M, namely
\begin{equation}
A^{M}(i,j) \defeq \frac{1}{f N^{2}} \times \left( \pi^{M}(i) - \pi^{M}(j)  + N \epsilon_{M}(i,j) \right)
\end{equation}
where $ \pi^{M}(i)$ is the position occupied by the place field of neuron $i$ in environment $\pi^{M}$, and $\epsilon_{M} \in \left\{ -1,0,1 \right\}$ ensures periodic boundary conditions.
\end{itemize}

Two simulations, one for A and one for B, are conducted. Parameters are carefully chosen such that the \textit{clump} phase is maintained, the bump thoroughly explores the environment, no spontaneous transitions occur. In other words, the same system is sampled in one of the two maps during the whole simulation; this mimicks the fact that the rodent explores a single environment in \cite{Jezek11}. Two \textit{reference sessions} are defined using the first half (5,000 steps) of each simulation, and a \textit{test session} is constructed by concatening the second halves, for a total of 10,000 total time steps. Parameters used in the following analysis are: $T=0.006$, $N=1000$, $w = 0.05$, $f=0.1$.

\subsubsection{Decoding Results}

As a measure of decoding precision we use the true positive rate (TPR), i.e. the overall fraction of correctly-classified neural pattern. 
\begin{equation}
\text{TPR} \defeq \frac{\text{\# correctly classified time steps}}{ \text{\# total time steps}}
\end{equation}
We obtain 
\begin{align}
\text{Ising model : } \ \text{TPR} &= 0.928 \\ \notag
\text{Independent model : } \ \text{TPR} &= 0.491
\end{align}

The difference between the use of independent and Ising model,  shown in Fig.~\ref{subsampling-plot}, is remarkable. The independent model, in which all couplings are set to zero, accounts only for the average firing rates of the cells. It shows no decoding capability at all, with a TPR equal to 0.49 (compatible with random guessing). This could be expected from the fact that the localized bump of activity, which represents position of the rat within the retrieved a map, moves along the entire environment during reference sessions. Hence the average activity of all cells is close to $f$ in both maps. The independent model, which only uses information on averages to decode the activity, is therefore unable to achieve useful discrimination.

Conversely, the Ising model exhibits an impressive performance in the decoding task. As shown in Fig.~\ref{sub-plot-is}, the time course of the likelihood difference $\Delta {\mathcal L}$ allows us to unambiguously decode the spatial representation as a function of time. This difference is also clear from the scatter plot of the likelihoods in the test session, which shows a well-separated pattern in the plane, contrary to the Independent model (Fig. \ref{subsampling-scatter}). Computation of true positive rates fully justifies the remarkable visual difference between the two models:

\begin{figure}[h]
\hspace*{\fill}%
\centering
  \subcaptionbox{Independent model\label{sub-plot-ind}}{\includegraphics[width=\linewidth]{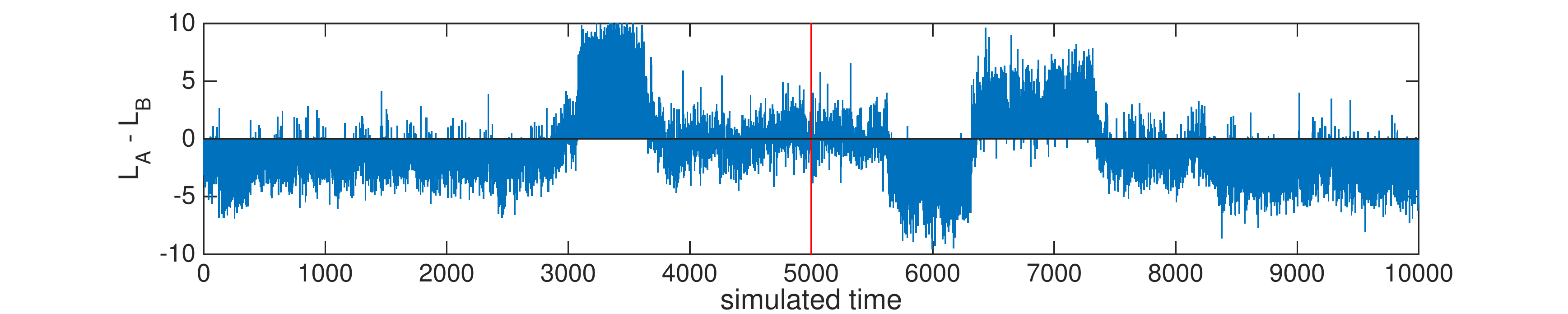}}\hfill%
  \centering
  \subcaptionbox{Ising model\label{sub-plot-is}}{\includegraphics[width=\linewidth]{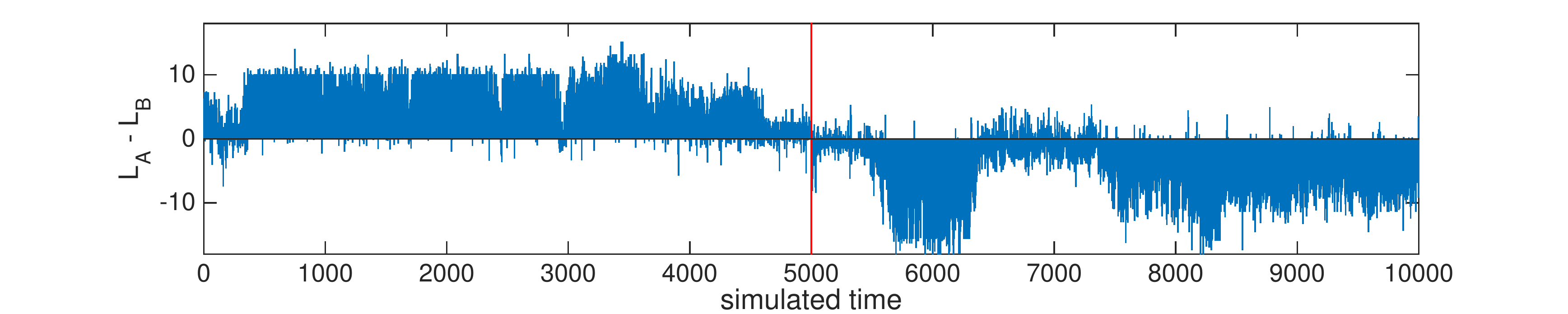}}%
  \hspace*{\fill}%
  \caption{Log-likelihood difference $\mathcal{L}_{A}(t) - \mathcal{L}_{B}(t)$ along the test session using independent model and Ising model on the montecarlo test session. The first half of the test session is sampled from environment \textbf{A}, the second half from environment \textbf{B}.}
\label{subsampling-plot}
\end{figure}

\begin{figure}[h]
\hspace*{\fill}%
  \subcaptionbox{Independent model\label{sub-scat-ind}}{\includegraphics[width=0.5\linewidth]{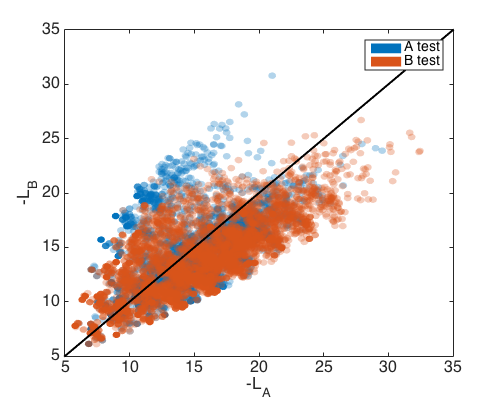}}\hfill%
  \subcaptionbox{Ising model\label{sub-scat-is}}{\includegraphics[width=0.5\linewidth]{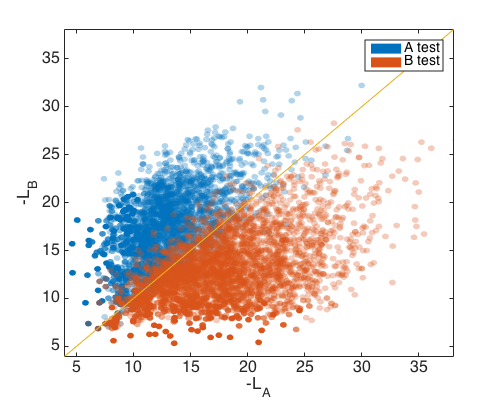}}%
  \hspace*{\fill}%
  \caption{Likelihoods scatters computed from the Independent (a) and Ising (b) models. Each dot represents the value of $-\mathcal{L}_{A}$ and $-\mathcal{L}_{B}$ for each neural configuration ${\bf s}^t$ during the Monte  Carlo test session.}
\label{subsampling-scatter}
\end{figure}

\subsubsection{Inferred vs. true couplings}

\begin{figure}[t]
\hspace*{\fill}%
  \subcaptionbox{Environment A\label{sub-scat-A}}{\includegraphics[width=0.4\linewidth]{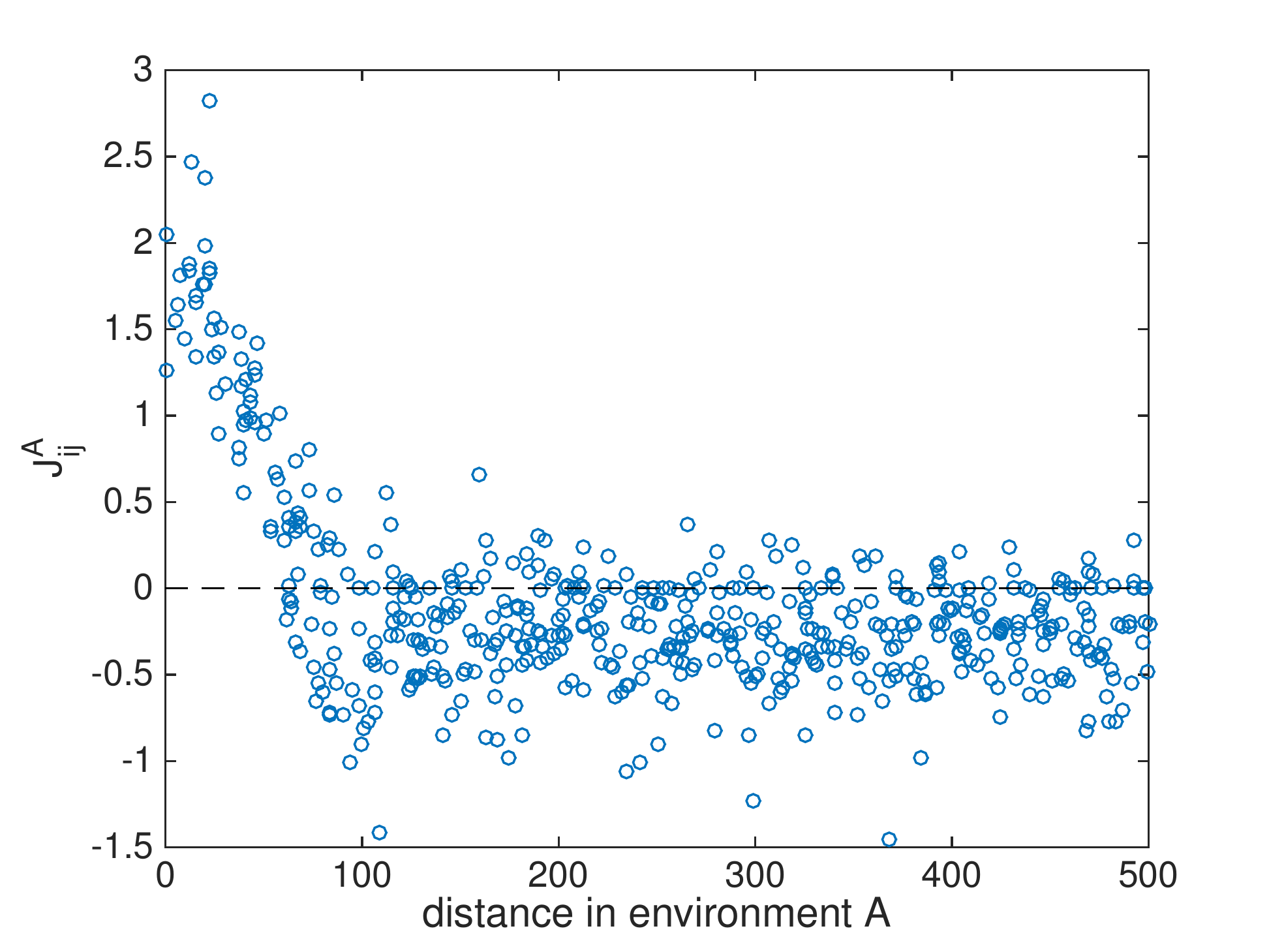}}\hfill%
  \subcaptionbox{Environment B\label{sub-scat-B}}{\includegraphics[width=0.4\linewidth]{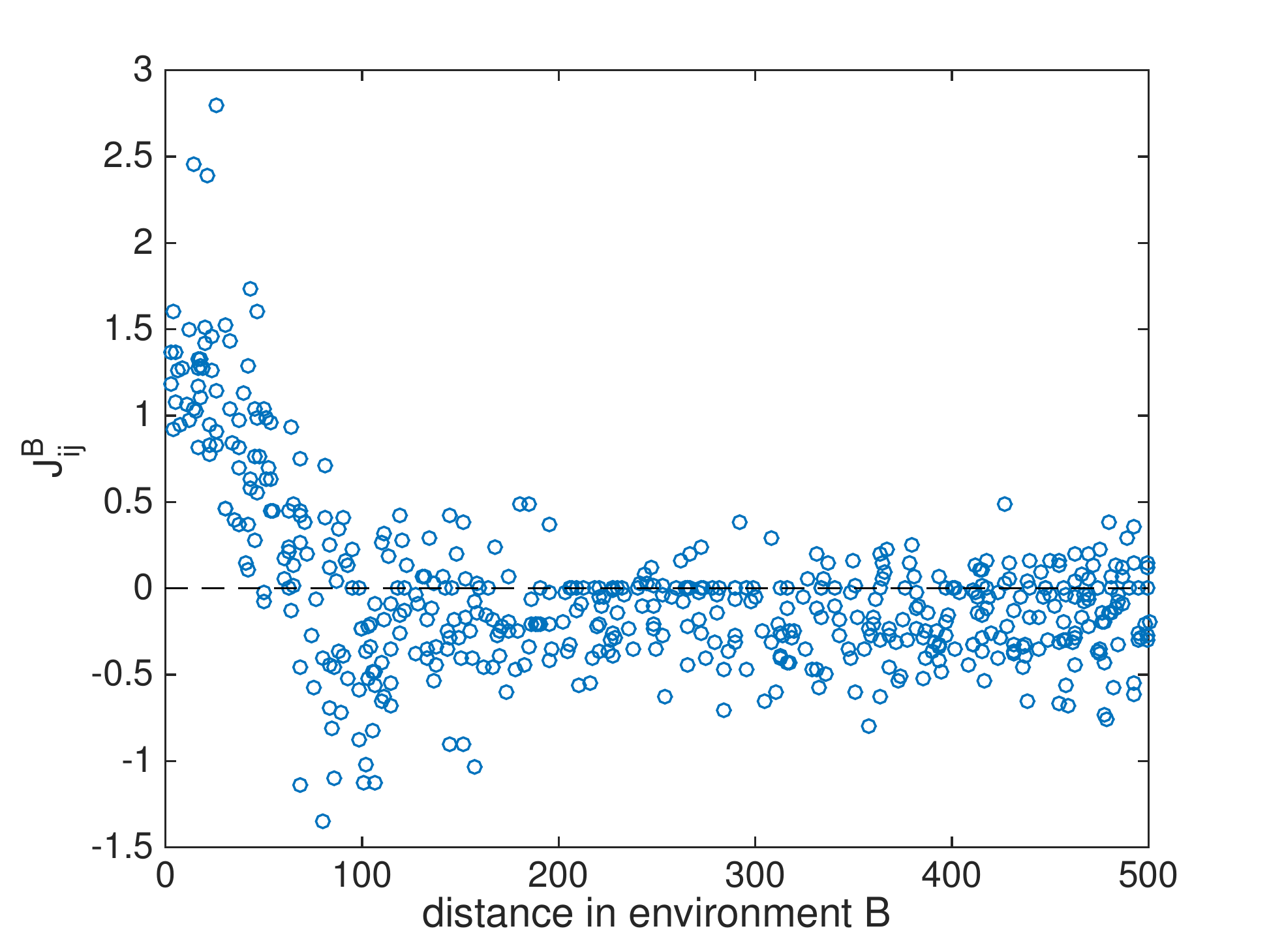}}%
  \hspace*{\fill}%
  \caption{Inferred coupling $J_{ij}$ vs. distance $\left | \pi^M(i) -  \pi^M(j) \right |$ between the place-field centers of the corresponding neurons in environment $M=A$ (left) and $M=B$ (right).}
\label{scatterJ}
\end{figure}

\begin{figure}
\hspace*{\fill}%
  \subcaptionbox{Environment A\label{sub-scat-A}}{\includegraphics[width=0.4\linewidth]{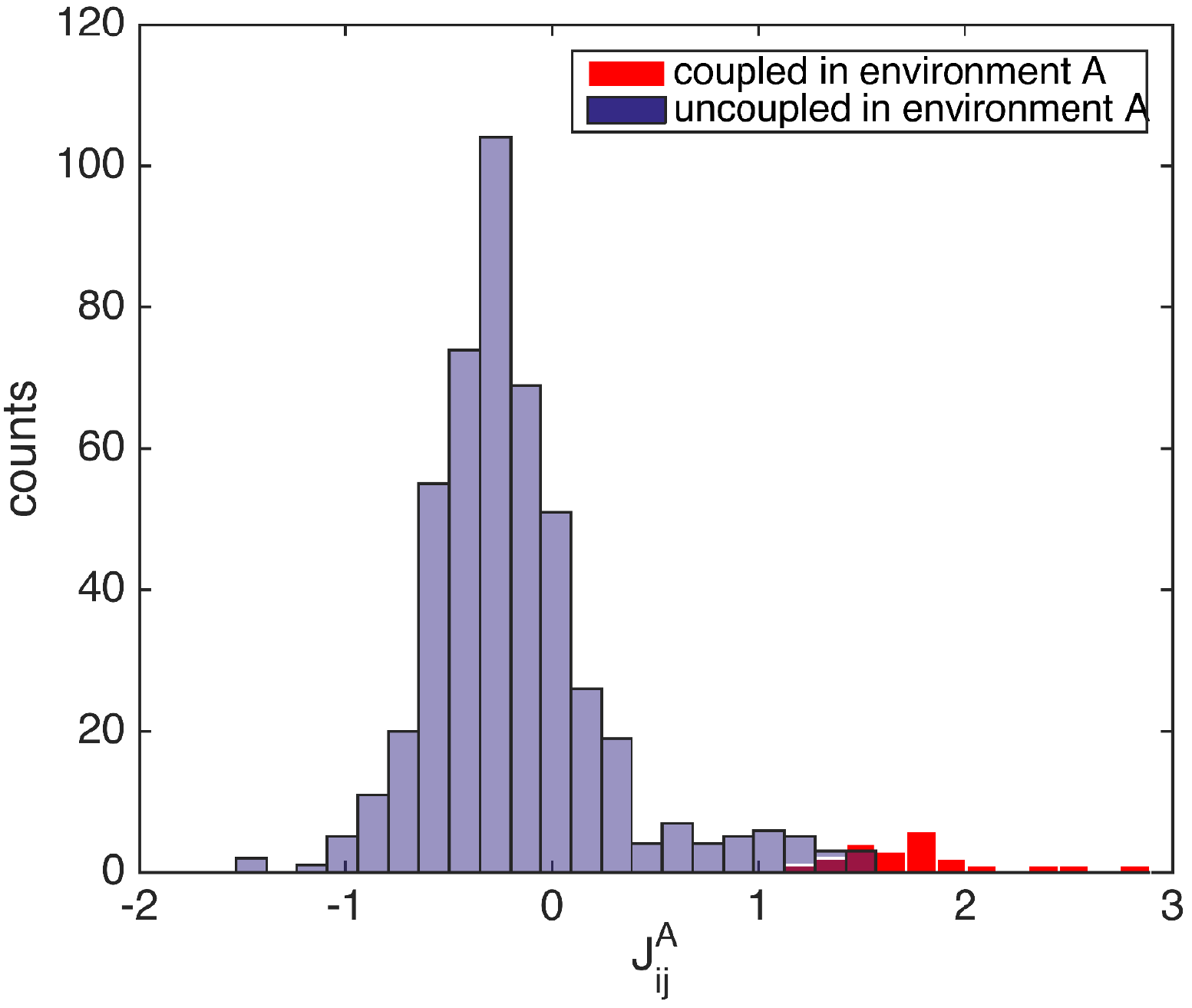}}\hfill%
  \subcaptionbox{Environment B\label{sub-scat-B}}{\includegraphics[width=0.4\linewidth]{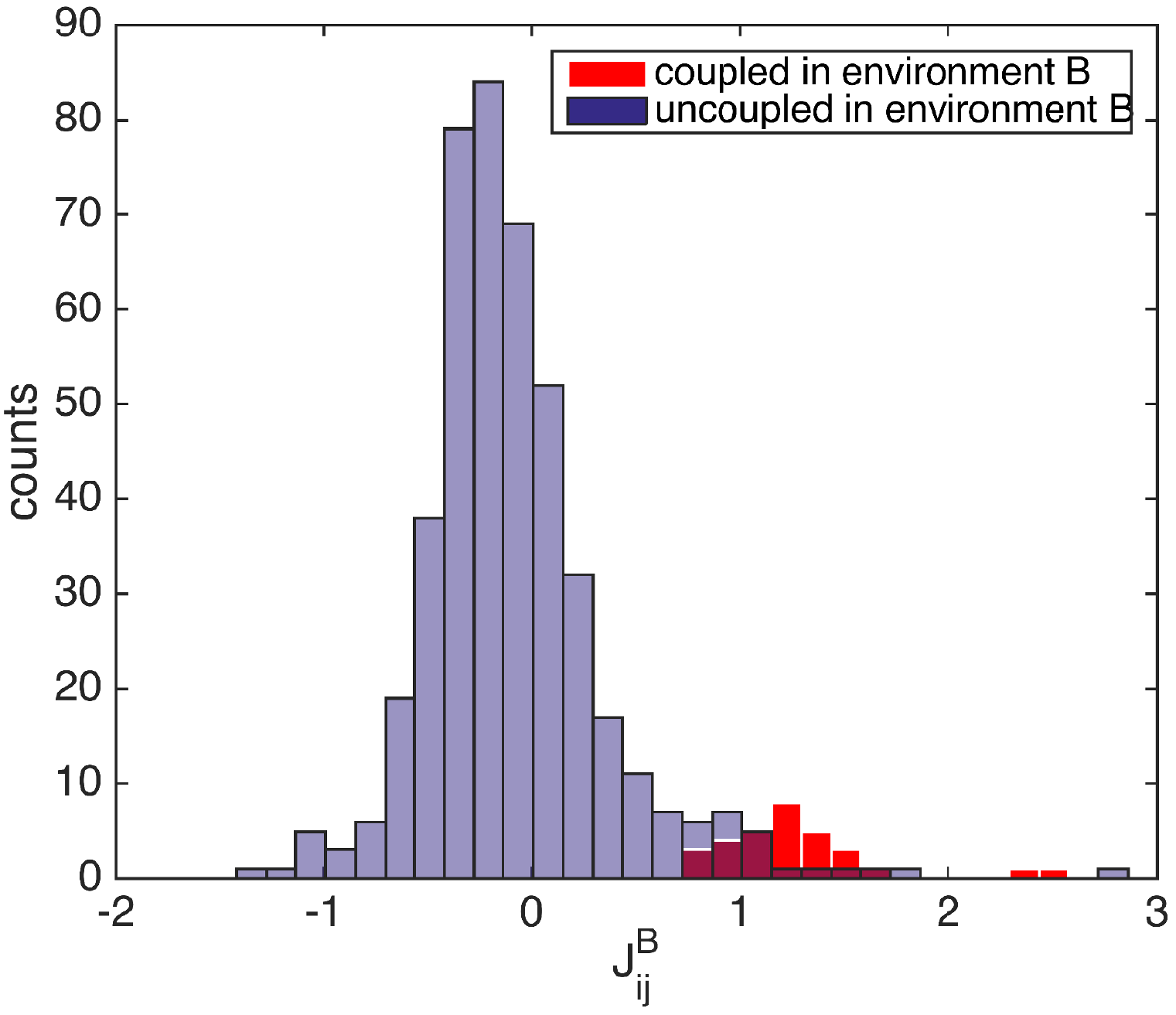}}%
  \hspace*{\fill}%
  \caption{Relationship between true couplings and inferred couplings. In purple, historgram of inferred couplings. In red, inferred couplings corresponding to truly connected neurons in the environment.}
\label{histoJ}
\end{figure}

The application of inference routines to a simulated neural network allows us to investigate the relationship between functional couplings, i.e. the inferred $J_{ij}$ in the inverse Ising model, and the real coupling strength, defined in eqn. (\ref{neuron}). We show in Fig.~\ref{scatterJ} the couplings inferred between the neurons as functions of the distances between their palce-field centers in each map. We observe that:
\begin{itemize}
\item Couplings decay very rapidly with the distance, on a typical scale compatible with both $w\,N$ and $f\,N$, and the width of the bump; Note that $w,f$ have similar values in the simulations. At long distances, couplings are independent of distance, and equal to a negative value. The presence of many long-range inhibitory couplings, clearly visible in the histrograms of Fig.~\ref{histoJ}, is a natural consequence of constraint (\ref{constraint1}) on the level of activity.
\item The magnitude of coupling at small distances, $\sim 2-3$ in Fig.~\ref{scatterJ}, is much larger than the one of the 'true' couplings in the model, equal to $J^0=\frac 1{T\,N}=0.167$. This suggest that the inferred couplings are {\em effective}, and would coincide with the true couplings only in the limit of perfect spatial sampling ($N_{sam}=N$).  
\end{itemize}
To better understand the value of the inferred couplings, let us compute the statistical moments of the neurons. As explained above in the description of the independent-cell models, all neurons have the same average activity in both environments:
\begin{equation}
p_i^A=p_i^B= f\ , \quad \forall i\ .
\end{equation}
We can also estimate easily the joint probability that two neurons are active.  In the large-$N$ limit, the true couplings vanish as they scale as $1/N$. Hence, spins become two-by-two independent in a ground state of the Hamiltonian, that is, when the bump is centered around a given position $x$. Conditioned to $x$, and defining the position of the place-field center of cell $i$ in map $M$ through
\begin{equation}
x_i =\frac{\pi^M(i)}N \ ,
\end{equation}
we have
\begin{equation}
\langle s_i\rangle_x = \rho\big(x_i-x\big), \ \langle s_j\rangle_x = \rho\big(x_j-x\big),\  \langle s_i s_j\rangle_x -\langle s_i\rangle_x \langle s_j\rangle_x \sim \frac 1N\ ,
\end{equation}
where $\rho$ is implicitly centered in 0 in the above expression.

However, we have to average over the position $x$ of the bump that moves across the environment (Fig.~\ref{montecarlo}). Doing so, we obtain the pairwise activity, see eqn (36) and Fig.~12 in \cite{Monasson13}:
\begin{equation}
p_{ij}^M= \int dx \, \rho\big(x_i+x\big)\, \rho\big(x_j+x\big)\quad \forall i,j\ .
\end{equation}
This effective matrix of pairwise activities threfore depends on the map, which explains why the Ising model, contrary to the Independent model, is map-specific and can efficiently decode the representation. However, the effective correlation between neurons, $p_{ij}^M-f^2$, does not scale as $\frac 1N$: the Ising couplings are thus effective interactions, not simply related to the true couplings in the model. We expect this statement to hold also for the functional couplings inferred from real recordings and their physicological, synaptic counterparts.

\subsection{Analysis of multi-electrode recordings in CA1}\label{secdata2}

The inference routines and the test-session validation described above are directly applicable to micro array recordings of neural activity in vivo. As previously introduced, one can record the brain area activity in a collection of stable memory states, build models from these activties, and use them to decode  representations (neural states) in a successive test session.

A good testing ground for this analysis is spatial memory in the rat hippocampus. Once environments have been memorized by the animal, it is relatively easy to collect samples of one memory state by letting the rat explore the corresponding environment.  In a recent experiment, conducted by Jezek et al.  \cite{Jezek11}, environmental conditions (light cues) are abruptly changed to trigger instabilities in the evoked spatial maps in the test session. By decoding which representation is being expressed during the test session as a function of time, we can investigate the response and fast dynamics of the memory state in the hippocampal network. 

\begin{figure}
\hspace*{\fill}%
\begin{center}
  \subcaptionbox{Independent model\label{ca1-plot-ind}}{\includegraphics[width=.95\linewidth]{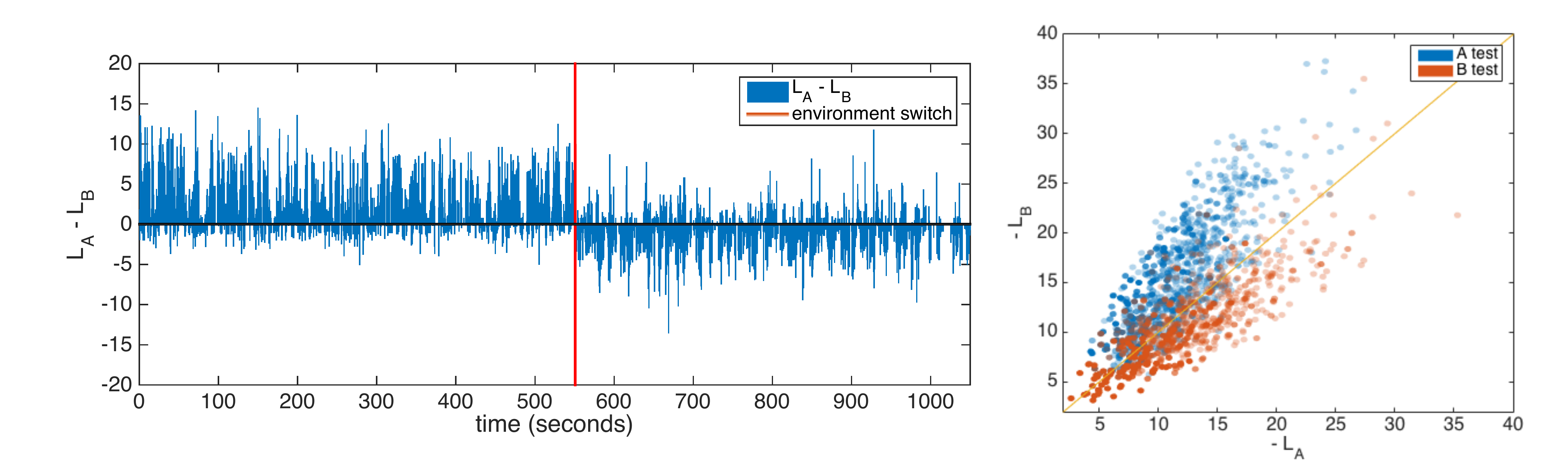}}\hfill%
  \subcaptionbox{Ising model\label{ca1-plot-is}}{\includegraphics[width=.95\linewidth]{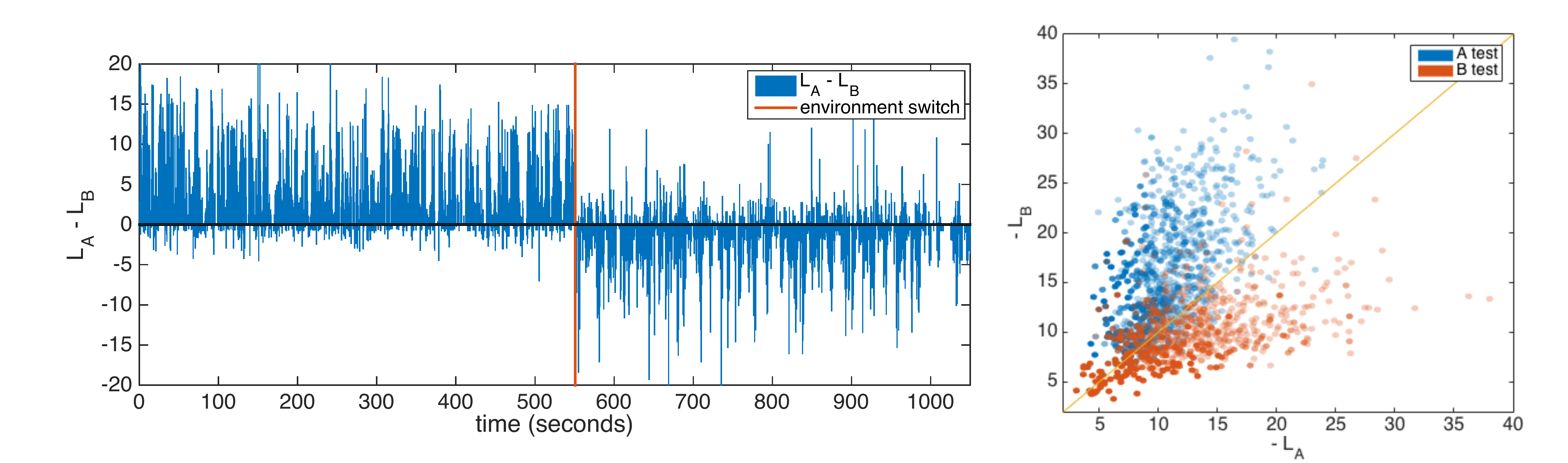}}%
  \hspace*{\fill}%
  \end{center}
  \caption{Log-likelihood difference $\mathcal{L}_{A}(t) - \mathcal{L}_{B}(t)$ along the test session using Ising and independent decoder on CA1 hippocampal data. The environmental conditions are abruptly changed from A to B in correspondence to the red line.}
\label{ca1-teleportation}
\end{figure}

The Ising inference method has been used in this context to decode which map, denominated A or B, is retrieved during test sessions with unknown environmental conditions. A test session recorded from hippocampal CA1 containing an environment-switch (teleportation) event is shown in Fig.~\ref{ca1-teleportation}. Contrary to the decoding analysis performed on the subsampled theoretical model, the Iindependent-cell model shows good performance in decoding task on real recordings:
\begin{align}
\text{Ising model: } \ \text{TPR} &= 0.85 \\ \notag
\text{Independent model: } \ \text{TPR} &= 0.81 \notag
\end{align}
One possible reason is that our theoretical model, designed to describe the memory storage and retrieval functions of the hippocampal network, does not account for the anatomical context of the region. In the mammalian brain, the hippocampus plays a role in a complex neural circuitry that involves inputs and outputs to other brain areas, such as medio entorhinal cortex (MEC) and lateral entorhinal cortex (LEC). The sensory input, conveyed by LEC and MEC, can dramatically change the firing properties of hippocampal place cells with respect to external environmental conditions, even to the point of silencing neurons in all but one environments. The mean neural activity could therefore carry enough information to achieve useful discrimination between the explored environments and the corresponding recalled memory states, while it is identical in all maps and therefore useless for decoding in our model. The properties of remapping of place cells activities in different representations have been extensively studied in the neuroscience literature, and different features have been reported on different hippocampal sub-regions \cite{fyhn2007hippocampal}. For a more detailed analysis and comparison of inference methods applied to hippocampal data see \cite{posani2017functional, posani2017position}.

\section{Representations of Data in Machine Learning}

\subsection{Introduction}

We start by the definition of a data representation. Suppose we are given a set of $P$ data samples $\bf{x^{(1)}},\bf{x^{(2)}},...\bf{x^{(P)}}$ of a $N$-dimensional random variable $\bf{X}$ having joint density $P(\bf{X})$. A data transformation is a deterministic transformation from the multidimensional vector space of data into another one: 
\begin{equation}
F: \bf{x} \in \mathbb{R}^N \rightarrow {\bf x'} = F(\bf{x}) \in \mathbb{R}^M \ ,
\end{equation}
where $M$ can be larger or smaller than $N$. In general, $F$ is assumed to be differentiable, but is not necessarily invertible. We say that the random vector $\bf{X}'$ is a representation of the original random vector $\bf{X}$. Changing the representation of a random variable can be often extremely helpful in data science because: i) it allows for better visualization and understanding of the process that generated the data; ii) the performance of machine-learning algorithm, such as classification or clustering methods heavily depends on the choice of representation used. 

Although it is not obvious that a given representation is good, it is clear that many, many representations are useless: if $F({\bf X}) = 0, \forall {\bf X}$, then ${\bf X'}$ is a trivial random variable, and does not carry any information about ${\bf X}$. More generally, it is clear that any transformation $F$ that does not vary strongly across the support of ${\bf X}$ is of little use. On the opposite, $F=Id$ is not of much use either, since the properties of the data distribution have not changed. Typically, a good data representation ${\bf X}'$ must have helpful properties that ${\bf X}$ does not have, such as low dimensionality, independence between components or sparse values, while carrying information on the original random vector ${\bf X}$. Thus, the transformation $F$ must depend on $P({\bf X})$ and should be learnt. Once learnt, a data representation can often shed light on how the data was generated: one can find so-called 'features', i.e. frequent collective modes of variation in the data, find a partition into classes, discover outliers... 

Moreover, a good data representation can significantly improve the performance of subsequent machine learning tasks, by retaining only useful information about the data sample. For instance, in so-called deep neural network, one learns a sequence of data transformations, e.g. to predict label from an image. By using non-linearities and so-called pooling architectures, the learnt intermediate representations of the data can become invariant, e.g. w.r.t. noise, shifts, rotations... hence learn quicker \cite{mallat2016understanding}. Deep neural networks have brought remarkable breakthrough in many areas, such as visual and speech recognition, natural language processing,... \cite{Bengio2013,LeCun2015}

We now illustrate these concepts with two examples of great relevance in applications. 

\subsection{Example 1: Dimensionality reduction} \label{secpca0}
One important subclass of data transformation are Dimensionality Reduction transformations. One aims at compressing a random vector ${\bf X}$ of typically high dimension $N$, into a smaller random vector ${\bf X'}$ of dimension $M<N$, e.g. $M=2$ or $3$, while keeping as much information as possible about ${\bf X}$. Such compression is motivated by the fact that data very often lie in or close to a subspace of much lower dimension than $N$. This is the so-called 'manifold hypothesis'. Indeed, consider for instance a data set constituted by pictures of a person's face, taken in many different positions; each picture is made of, say, $1000\times 1000$ pixels. It is clear that this data set is a very small subset of all possible $1000\times1000$ colored pictures, which define a $3 \,10^6$--dimensional vector. The reason is that, for a given face, there are only $\sim 50$ varying degrees of freedom (the position of all muscles), a very small number compared to $10^6$ \cite{LecunCDF}. Hence, all data points lie in a (non-linear) manifold, of very low dimension $M$ compared to $N$. More generally, the variability in the data often comes from a small number of explanatory latent factors that affect all components, and we would like to recover them.

%\begin{figure}
%\begin{subfigure}{.45 \textwidth} \label{manifold_hypothesis}
%\includegraphics[scale=0.5]{img2/manifold_hypothesis.jpg}
%\end{subfigure}
%\begin{subfigure}{.45 \textwidth}
%\label{manifold_hypothesis2}
%\includegraphics[scale=0.5]{img2/manifold_hypothesis2.jpg}
%\end{subfigure}
%\caption{(a) Pictures of a person with various facial expressions. They lie in a very low dimensional manifold of the vector space of pictures with $1000\times 1000$ pixels. (b) Example of complex 2D manifold embedded in a 3D vector space.}
%\end{figure}

\begin{figure}
 \begin{subfigure}{.33 \textwidth}
 \label{MNIST_samples}
 \includegraphics[scale=0.38]{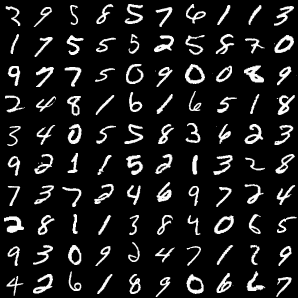}
  \end{subfigure}
\begin{subfigure}{.33 \textwidth}
\includegraphics[scale=0.38]{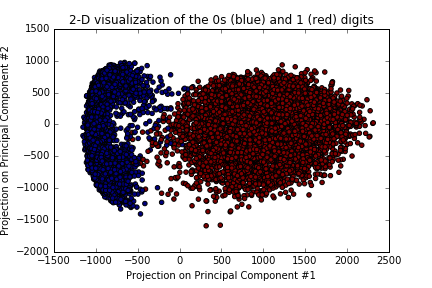}
 \end{subfigure}
 \begin{subfigure}{.33 \textwidth}
 \label{PCA_MNIST2}
 \includegraphics[scale=0.38]{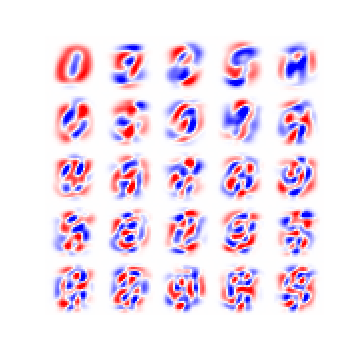}
  \end{subfigure}
\caption{(a) Some MNIST data samples. (b) A 2-dimensional PCA representation of the MNIST handwritten digits data set. Each point is a different image with $x$ and $y$ coordinates being the value of the first and second components of the representation. Here, only the digits 0's (blue) and 1's (red) are represented. (c) Visualization of the weight matrix $W$. Each image is a principal component vector $W_{i,.}$; blue (resp. red) pixels denote large positive (resp. negative) values. the PCA representation is obtained by computing the set of overlaps between an image and each principal component vector}
\label{PCA_MNIST}
\end{figure}

In all generality, we do not have good and general methods to learn functions that turns an image into this kind of 'muscle positions' representation. Some simpler dimensionality reductions can nonetheless be learnt and be extremely useful. For instance, dimensionality reduction can be obtained through a simple linear transformation: 
\begin{equation}
{\bf X'} = W \, {\bf X}\ .
\end{equation}
where the weight $W$ is a $M \times N$ rectangular matrix that must be trained on the data in order to retain as much information as possible from ${\bf X}$. An interesting choice of matrix $W$ is obtained by the Principal Components Analysis (PCA) algorithm: the rows $W_{i,.}$ are the eigenvectors corresponding to the $i$'th largest eigenvalues of the empirical data covariance matrix $C_{ij} =\langle X_i X_j \rangle-\langle  X_i \rangle\langle X_j \rangle$, where the average is computed over the data; this choice will be justified in Section \ref{secpca}. Such transformation mainly serve two purposes. The first one is to provide a better understanding of the data by visualizing it: one computes a 2 or 3 dimensional-representation of the data; then each data point is represented in a 2 or 3D space. For example, one can compute the 2D PCA representation $28 \times 28$ images of digits from the MNIST handrwitten digits dataset, vectorized as $784$ dimensional vectors, see Fig.~\ref{PCA_MNIST}; the scatter plot shows two distinct clusters, corresponding to two digit types (0s and 1s). A more interesting illustration is the interpretation of molecular dynamics simulation of complex systems, made of many strongly interacting and heterogeneous microscopic components. Observing the dynamics of such systems, e.g. a protein described at the atomic level, amounts in practice to look at thousands of correlated time series. Principal component analysis offer low-dimensional projections of these time traces, and allows one to visualize collective motions underlying the evolution of the system, see \cite{muellerstein06} for a recent review on applications to biomolecules, including nucleic acids and proteins.

The second purpose of dimensionality reduction is to overcome the so-called curse of dimensionality. In very high dimensional spaces, most datasets sample only very sparsely the vector space $\mathbb{R}^N$.  Consider for instance the following supervised learning problem. We are given a training data basis of $10,000$ $100 \times 100$ grayscale (normalized between 0 and 1) images of cats and dogs, with binary labels attached, and we want to train a parametric model to classify whether images are cats or dogs. At this point, it is useful to think that this classification task is essentially an interpolation problem: there exist a mathematical function $\theta: {\bf X} \rightarrow y \in \{0,1\}$ that assigns $0$ to cats and $1$ to dogs. We observe pairs of values $ \left( {\bf X}^i, y^i =\theta({\bf X}^i) \right)$, with $i=1...10,000$, and want to interpolate the values of $\theta$ for new test images. This interpolation problem would be trivial if the input space was densely sampled, \textit{e.g.} if for any point in $\mathbb{R}^N$ there would be a training data point at distance $\leq \epsilon$. In practice, it is impossible because the latter condition requires about $\epsilon^{-N}$ data points, which is out-of-reach when $N$ is large. 

One possible way-out is to first learn a new data representation of lower dimension, $\bf{x'} = F(x)$, e.g. using PCA, and then train a classification model of the form: $y = \theta(\bf{x'})$. If the low dimensional representation keeps relevant information about the nature of the image, then learning can be performed. One popular application of PCA for supervised learning is the 'eigenface' face recognition algorithm. A PCA representation is trained on a data set of faces, before applying supervised learning \cite{turk1991face}. The eigenface algorithm is considered among the first successful face recognition algorithms.

\subsection{Example 2: Extracting latent features from data}
\begin{figure}
\centering
\includegraphics[scale=0.25]{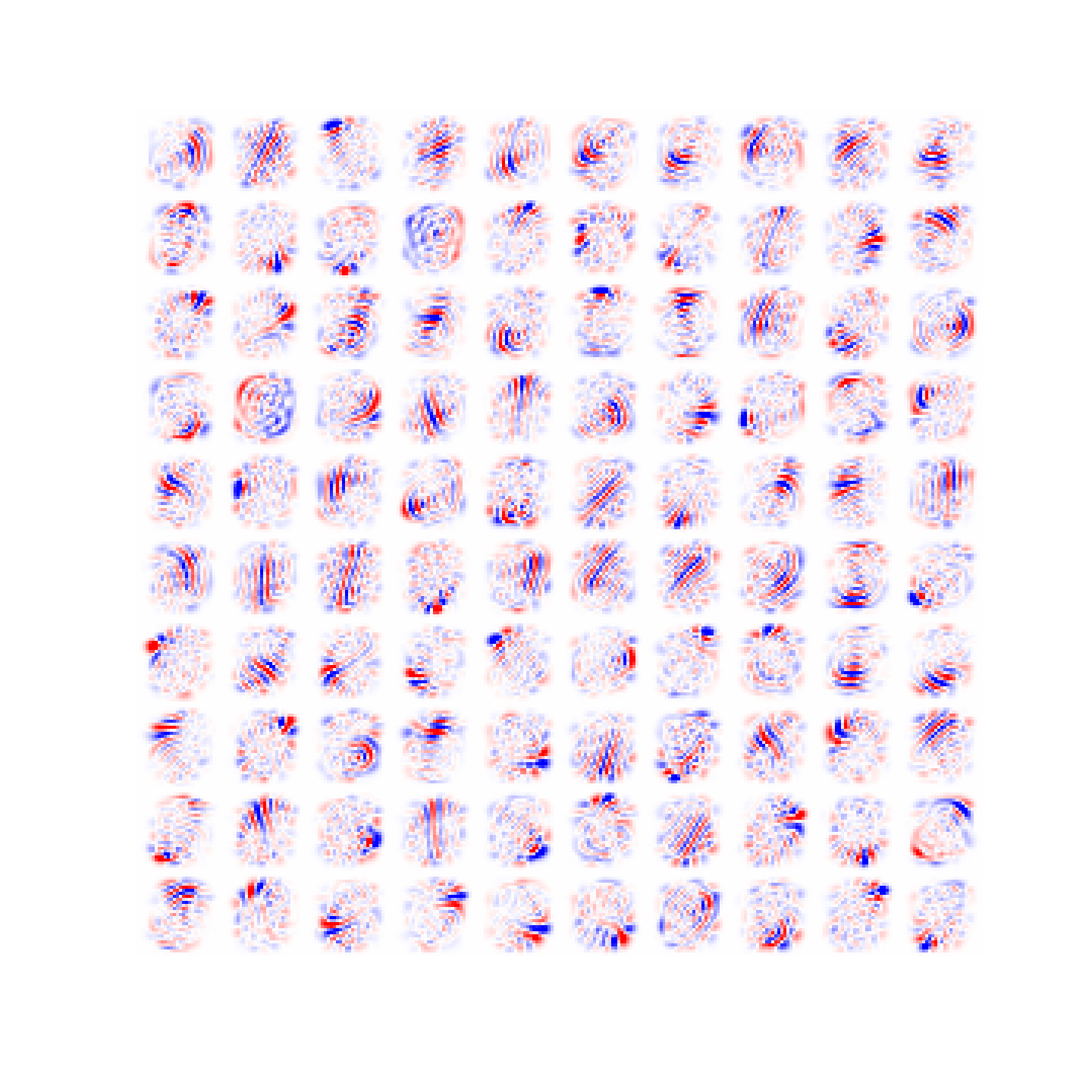}
\vskip -1cm
\caption{Features learnt by Independant Component Analysis on MNIST}
\label{ICA_MNIST}
\end{figure}

The variability in real-world data, such as images, can often be decomposed into a set of largely independent modes of variation. For instance, two faces are different because some of their parts are different: nose, ears, lips... At a lower level of description, an image can contain or not an edge at a given location, or at some angle or scale, and two different images have different set of activated edges. Extracting these so-called 'features' is of particular interest for machine learning, in particular for classification, because the decision function $y = \theta(X)$ that must be learnt may be expressed more easily as a function of these 'features' $X'$ than from the raw pixels $X$. For instance, one could achieve better results by expressing $\theta(X')$ as a linear function of $X'$, instead of a higher order polynomial of $X$. Moreover, the learnt representations have interesting statistical properties, such as low statistical dependence between modes, invariance with respect to irrelevant perturbations of the data such as corruption by noise... that can be used for denoising. Some notable algorithms for unsupervised feature extraction are Independent Component Analysis (ICA) \cite{hyvarinen2004independent}, sparse autoencoders \cite{ng2011sparse}, and sparse dictionary learning \cite{olshausen1996emergence}. We display in Fig.~\ref{ICA_MNIST} the features learnt by ICA applied to the MNIST digits data set. The features learnt correspond to individual handwritten strokes, unlike PCA where the principal component do not have a simple interpretation. Interestingly, the features found by sparse dictionary learning applied to natural images dataset qualitatively match very well the receptive fields of neurons in the visual cortex of mammalians, such as in monkey \cite{ringach2002orientation,zylberberg2011sparse}. Feature extraction carried out in the brain bear strong analogies with machine-learning procedures. \cite{poggio16}.

\section{Low-Dimensional Representations: Principal Component Analysis (PCA)}
\label{secpca}

In this section, we focus on the PCA transformation introduced in Section \ref{secpca0}. To cast PCA in a Bayesian framework, we start with a basic reminder about Bayes's approach to inference.

\subsection{Mathematical reminder: Bayesian inference}
\begin{figure}
  \begin{minipage}[c]{0.5\textwidth}
    \includegraphics[width=\textwidth]{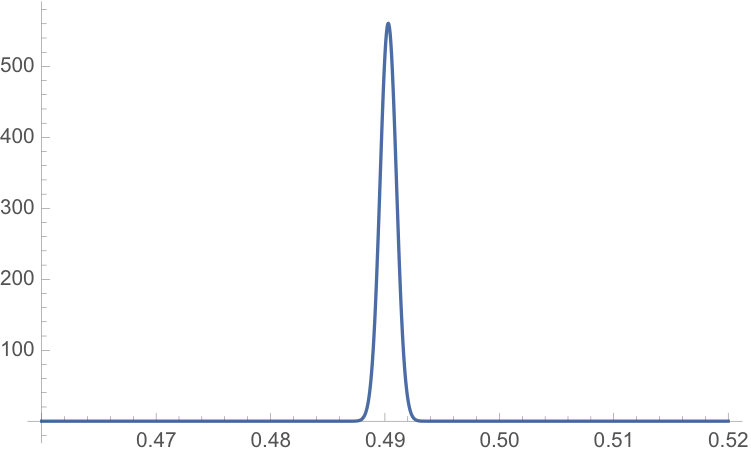}
  \end{minipage}\hfill
  \begin{minipage}[c]{0.45\textwidth}
    \caption{The posterior probability density of the female birth rate $\tau$ for Laplace's birth rate problem according to eqn (\ref{lapla2}). Notice that  $p(\tau = 0.5)$ is very small, but non zero.  } 
    \label{laplace_inference}
  \end{minipage}
\end{figure}

We observe a data sample $\sigma$,  and would like to fit these data with a model parametrized by some variable $\tau$. We assume that both $\sigma$ and $\tau$ are random variables, with a joint distribution $p(\sigma,\tau)$. According to the definition of conditional probabilities, we may write
\begin{equation}\label{bayes1}
 p(\sigma,\tau) = p(\sigma| \tau)\times p(\tau) =p( \tau | \sigma) \times p(\sigma) \ .
 \end{equation}
In the first equality, $p(\sigma| \tau)$  is the the probability of the data given the model parameters, also called \textit{likelihood} of the model parameters given the data.
The second term, $p(\tau)$, is the \textit{prior distribution} over model parameters. The expression can be rewritten using the posterior distribution of the parameters given the observations $p(\tau | \sigma)$ (which can be maximized, sampled from...), and the overall probability $p(\sigma)$ of the data to be generated by the class of models uner consideration. This posterior distribution is given by Bayes formula:
\begin{equation}
p(\tau| \sigma) = \frac{p(\sigma | \tau) p(\tau)}{p(\sigma)}\ ,
\end{equation}
which is simply derived from eqn (\ref{bayes1}).
One historical application of the Bayesian inference formula is Laplace's statistical proof that boys and girls have different birth rates. Laplace had access to the number of boys and girls born in Paris between 1745 and 1770: $\sigma =245,945$ girls out of $P= 245,945 + 251,527 = 497,472$ babies born during this time period. Although the numbers of male and female births are different, it is not possible to know a priori whether the discrepancy comes from a statistical fluctuation or from a systematic difference in birth rates. Laplace assumed that each birth is a realization of an independent and identically distributed random variable, giving a girl with probability $\tau$ and a boy with probability $1-\tau$. Under these basic assumption, $\sigma$ follows a binomial distribution $\mathcal{B}(P,\tau)$, with a likelihood:
\begin{equation}\label{lapla1}
p(\sigma | \tau) = \left( \begin{array}{rr} P\\ \sigma
\end{array} \right) \tau^\sigma (1-\tau)^{P-\sigma} \ .
\end{equation}
Assuming a uniform density prior over $\tau\in[0;1]$, $p(\tau) = 1$, the posterior distribution reads 
\begin{equation}\label{lapla2}
p(\tau |\sigma) = C\; \tau^\sigma (1-\tau)^{P-\sigma} \ ,
\end{equation} 
where $C$ is a normalization constant, and is shown in Fig.~\ref{laplace_inference}. It is then easy to calculate the mean value and the standard deviation of $\tau$ with the posterior distribution, with the results Mean$( \tau )= 0.490291$ and Std$(\tau)=0.007117$. The probability that $\tau$ is actually larger or equal to $\frac 12$ is given by the integral of $p(\tau|\sigma)$ over the $\tau\in[\frac 12;1]$ interval, and is approximately equal to $ 10^{-42}$. This extremely small value makes it very unlikely that the discrepancies between the large numbers of female and male births is due to a pure statistical fluctuation.

\subsection{Multivariate Gaussian variables}\label{secprob}

A popular example of parametric model is the multivariate Gaussian distribution, used to model sets of continuous random variables exhibiting correlations. Hereafter, we will assume that all random variables have zero mean for the sake of simplicity. Given a vector of $N$ variables $\boldsymbol\sigma = (\sigma_1, \sigma_2, ... , \sigma_N )$, we write:
\begin{equation}
\rho(\boldsymbol\sigma | \boldsymbol\tau) = \frac{\sqrt{\det \boldsymbol\tau}}{ (2\pi)^{\frac{N}{2}}} \exp \left( - \frac{1}{2} \boldsymbol\sigma^T \cdot \boldsymbol\tau \cdot\boldsymbol \sigma \right) \, 
\end{equation}
where $\tau$, called \textit{precision matrix} is a symmetric, positive definite matrix that encodes the inter-dependencies between variables. Its off-diagonal entries can be interpreted as (minus) the couplings between the variables. For instance, with $N=2$, $\tau_{12} >0$ means that the configurations $(a,b)$ and $(-a,-b)$ are more likely that the configurations $(-a,b)$ and $(a,-b)$ (when $a,b>0$), leading to a positive correlation between $\sigma_1$ and $\sigma_2$. Given a data set of $P$ samples $\sigma^{(1)},\sigma^{(2)},..\sigma^{(P)} $, one can compute analytically the maximum likelihood estimator of the precision matrix:
\begin{equation}
\tau^{MLE} = \argmaxB \sum_{s=1}^P \log \rho (\sigma^{(s)}|\tau )  \ .
\end{equation}
To solve for $\tau^{MLE}$, the gradient of the right hand side in the above equation reads
\begin{equation}
\frac{\partial}{\partial \tau_{ij}} \sum_{s=1}^P \log \rho (\sigma^{(s)}|\tau ) = - \frac{1}{2} \sum_{s=1}^P  \sigma_i^{(s)} \sigma_j^{(s)} + \frac{P}{2} \, (\tau^{-1})_{ji} \ .
\end{equation}
We recognize that first term is the empirical data covariance matrix, $C$. The gradient vanishes -and it is easy to show that this corresponds to a global maximum- when
\begin{equation}
\tau^{MLE} = C^{-1}\ .
\end{equation}
The inversion can be performed numerically as long as $P\ge N$; for $P < N$, the data covariance matrix is not full rank. However, finite sampling effects of order $\frac{1}{\sqrt{P}}$ in $C$ result in error on $\tau = C^{-1}$ of the order of $\sqrt{\frac{N}{P}}$. Thus, for large-dimensional data sets, i.e. when the ratio $N/P$ is of the order of unity, we expect inference to be plagued with errors. 

\subsection{Principal components as minimal models of interacting variables}
\begin{figure}
\begin{subfigure}{.45 \textwidth}
\includegraphics[scale=0.5]{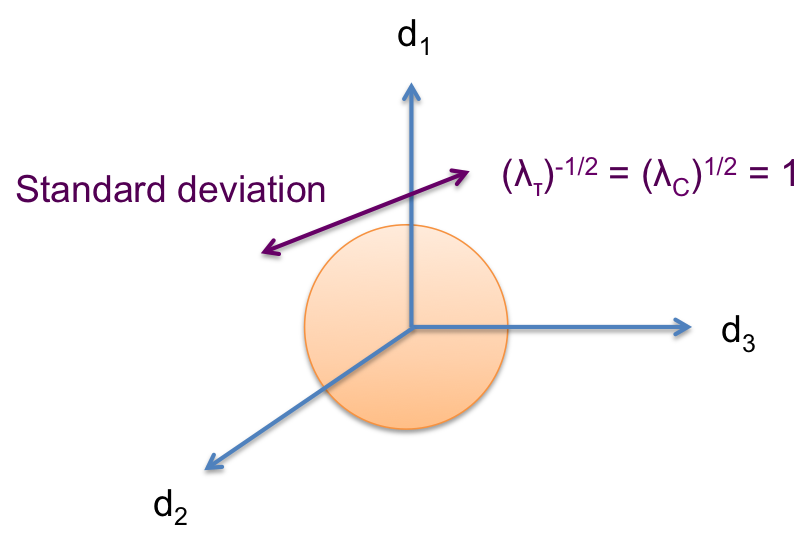}
\end{subfigure}
\begin{subfigure}{.45 \textwidth}
\includegraphics[scale=0.5]{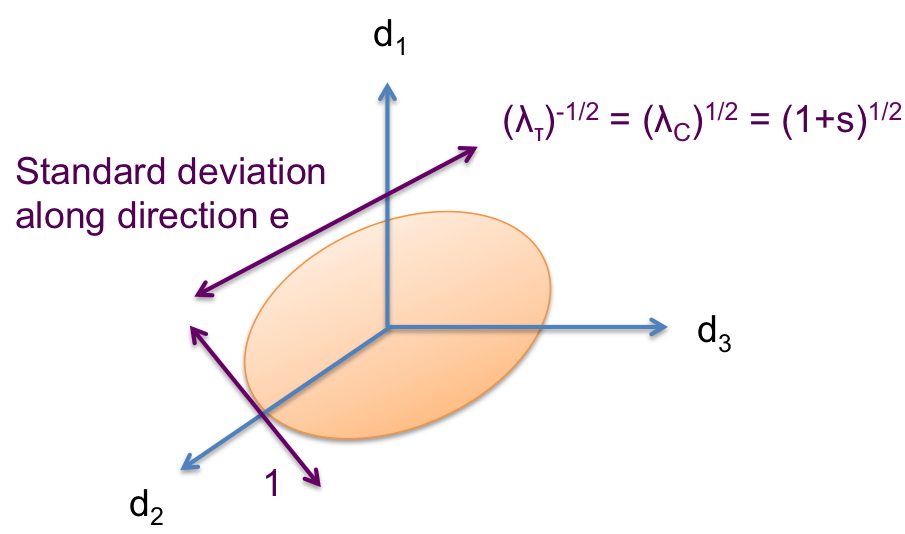}
\end{subfigure}
\caption{Probability density contours for (left) the null model ($\tau = Id$) and (right) the principal-component model of eqn (\ref{pcmodel1}).}
\label{null_model}
\end{figure}

The simplest model distribution over vector of random variables, each components being normalized to have zero mean and unit variance, is the independent one, which corresponds to $C =\tau= Id$. In this case, we have $p(\sigma | \tau) \propto \exp \left( - \frac{1}{2} \sum_i \sigma_i^2 \right)$, and the resulting distribution is isotropic, see Fig.~\ref{null_model}(left). A minimal non-trivial model is obtained by breaking this isotropy. We assume there exists a specific direction, denoted by $|e\langle$, in the $N$ dimensional space with a larger variance:
\begin{equation} \label{pcmodel1}
\begin{split}
\tau = Id - \frac{s}{1+s}\; |e\rangle\langle e|
\Longleftrightarrow C = \tau^{-1} = Id + s \; |e\rangle\langle e| \ ,
\end{split}
\end{equation}
where $s>0$. In this expression, $|e\rangle$, the \textit{principal component} can be interpreted as a \textit{collective mode} of variation of the data. Indeed, the random variable $\sigma_e = \sum_{i=1}^N e_i \sigma_i $ has variance $V_e = \langle e| C |e\rangle = 1+s$ larger than 1, whereas it would be 1 if the $\sigma_i$ were independent. The $\sigma_i$ variables correlate in a way that makes $\sigma_e$ have large variance, see Fig.~\ref{null_model}(right). 

\begin{figure}
\begin{subfigure}{.30 \textwidth}
 \label{eigen_null_infinite}
\includegraphics[scale=0.5]{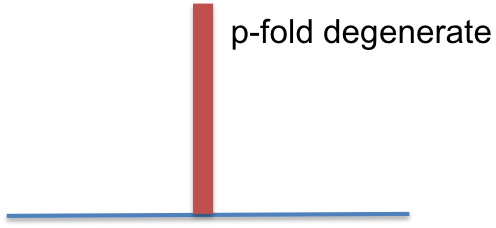}
\end{subfigure}
\vskip -.8cm \hskip 5cm
\begin{subfigure}{.30 \textwidth}
 \label{eigen_null_finite}
\includegraphics[scale=0.5]{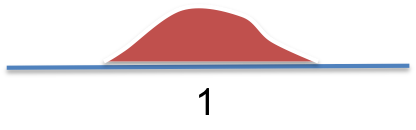}
\end{subfigure}
\vskip -1.55cm \hskip 10cm
\begin{subfigure}{.30 \textwidth}
\includegraphics[scale=0.5]{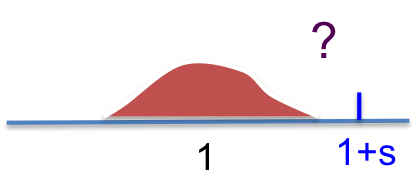}
\end{subfigure}
\caption{Distribution of the eigenvalues of the empirical covariance matrix for: (a) the null model with infinite sampling; (b) the null model with finite sampling; (c) the principal component model with finite sampling}
 \label{eigen_pca_finite}
\end{figure}

Given a data set, maximum likelihood estimation can be performed analytically to infer the principal component $|e\rangle$. The likelihood writes
\begin{equation}
\rho\big(\sigma^{(s)} \big| \; |e\rangle \big) = \frac{\sqrt{\det \boldsymbol\tau}}{(2 \pi)^{\frac{N}{2}}} \exp \left( -\frac{1}{2} \sum_{i,j} \sigma_i^{(s)} \tau_{ij}\, \sigma_j^{(s)}  \right) \ .
\end{equation}
The $|e\rangle$-dependent part of the log-likelihood is simply
\begin{equation} \label{maxL}
L =  \frac{s}{2(1+s)} \sum_{i,j} e_i\, e_j  \,\left( \sum_s \sigma_i^{(s)} \sigma_j^{(s)} \right) \ .
\end{equation}
Hence, the MLE for the direction $|e\rangle$ (assumed to be normalized) is the top eigenvector (with largest eigenvalue) of the empirical covariance matrix $C = \frac{1}{n} \sum_s \sigma_i^{(s)} \sigma_j^{(s)}$. Although the inference can be performed easily for any covariance matrix, we do not expect that the inferred vector is always  statistically significant, according to the discussion at the end of Section \ref{secprob}. For instance, even if the data are generated according to the null model $\tau=Id$, the empirical covariance matrix has a largest eigenvalue ($>1$) due to finite sampling (Fig.~\ref{eigen_pca_finite}). Similarly, if the data is generated according to the principal-component model but  $s$ is 'small' and $P$ is finite, the largest eigenvector of the empirical correlation matrix may be  far away from $|e\rangle$ (Fig.~\ref{eigen_pca_finite}). In the next section, we report analytical results derived using random matrix theory and statistical mechanics tools telling us when inference is possible.

\subsection{The retarded-learning phase transition}
\label{secretard}

We first study the empirical covariance matrix $C_{ij}= \frac{1}{n} \sum_s \sigma_i^{(s)} \sigma_j^{(s)}$, and its spectrum when the data are generated according to the null model $\tau = Id$. We are interested in particular in the empirical density probability of eigenvalues:
\begin{equation}
\rho(\lambda) = \frac 1N \sum_{\mu=1}^N \overline{\delta(\lambda_\mu - \lambda) }\ ,
\end{equation}
where $\{\lambda_1,...,\lambda_N\}$ is the set of eigenvalues of $C$. The overbar denotes the average over the realizations of the $N$ data samples (s). Note that, in the large $P,N$ limits with a fixed ratio $r \equiv \frac{N}{P}$, we expect that the spectrum attached to a random realization will coincide with the average spectrum  $\rho$ with high probability. 

\begin{figure}
\begin{center}
\includegraphics[scale=0.6]{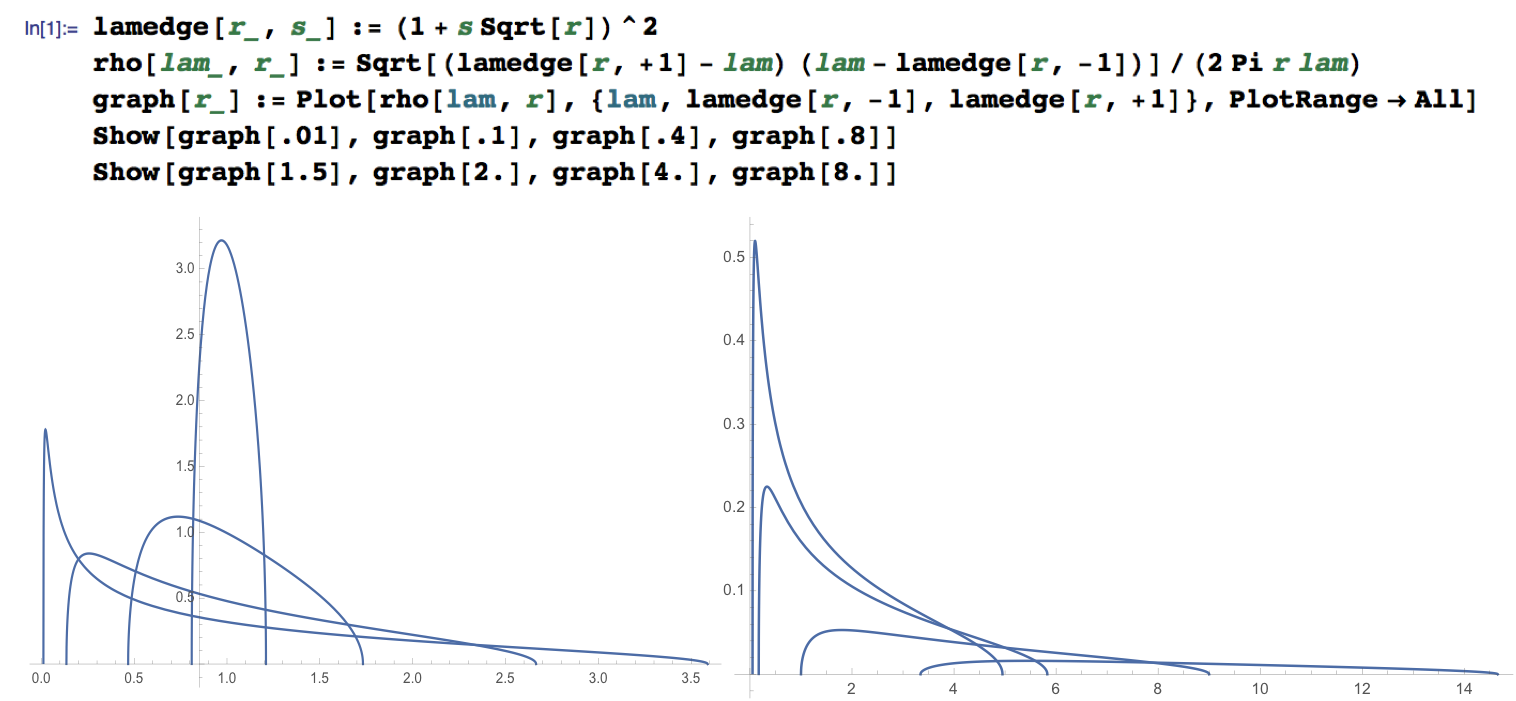}
\end{center}
\caption{The Marcenko-Pastur distribution of eigenvalues, $\rho(\lambda)$, for various values of the noise level $r$, reported in the Mathematica script on the top. Left: $r<1$. Right: $r>1$; the Dirac peak in $\lambda=0$ is omitted.}
\label{marcenko_pastur}
\end{figure}

The probability density of eigenvalues can be computed analytically using random matrix theory tools, in the limit case where the dimension $N$ and number of data points $P$ both go to infinity, and at fixed noise level $r$ \cite{marvcenko1967distribution}. The result is the so-called Marcenko-Pastur distribution:
\begin{equation}\label{mp}
\rho(\lambda)=\frac{\sqrt{(\lambda_+ - \lambda)(\lambda-\lambda_-)}}{2 \pi r \lambda} \ , \quad \text{with}\quad 
\lambda_{\pm} = \left( 1 \pm \sqrt{r} \right)^2 \ .
\end{equation}
The expression above is valid for $r<1$; for larger $r$, the covariance matrix is not full rank ,and there is also a Dirac peak of mass $1-\frac{1}{r}$ in $\lambda=0$. The distribution of eigenvalues is plotted in Fig.~\ref{marcenko_pastur} for various values of the noise level $r$. For very good sampling $r\to 0$, Wigner semi-circle law is recovered around $\lambda=1$, as the different entries of the correlation matrix becomes essentially uncorrelated. Interestingly, the spectrum density can be quite wide when $r$ is small: for instance, for $r=1$, eigenvalues can be as large as $\lambda=4$. As a consequence, these 'sampling noise' eigenvectors can screen away true principal components if $s$ is not too large. 

\begin{figure}
\begin{subfigure}{.30 \textwidth}
\includegraphics[scale=0.3]{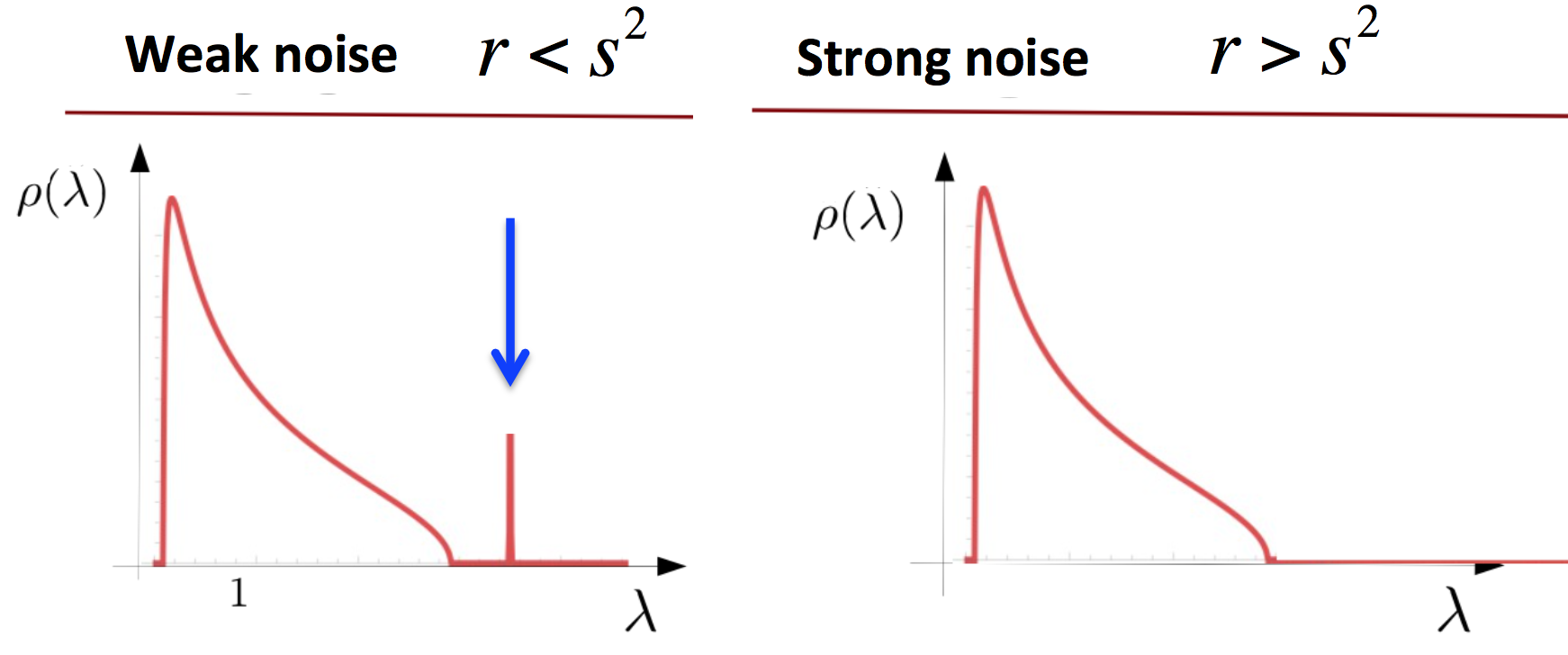}
\end{subfigure}
\hskip 5cm
\begin{subfigure}{.3 \textwidth}
\includegraphics[scale=0.4]{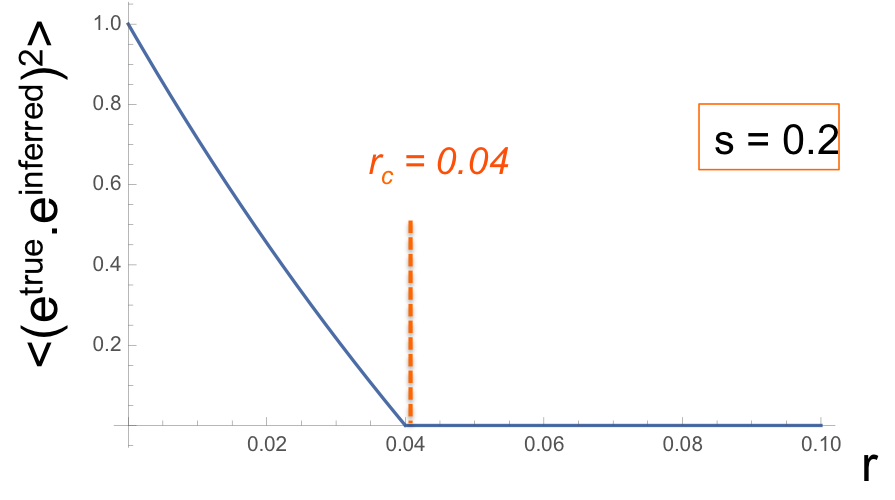}
\end{subfigure}
\caption{The retarded-learning phase transition. Left panels: spectrum of eigenvalues of the empirical correlation matrix in the principal-component model in the cases of weak noise (left, $r< s^2$) and of strong noise (right, $r>s^2$). Right panel: average squared overlap between the top components of the true and empirical correlation matrices as a function of the noise level $r$, for $s=0.2$.}
 \label{retarded_learning}
\end{figure}

The same computation as above can be carried out for the principal-component model with $s>0$, and shows the existence a phase transition, see Fig.~\ref{retarded_learning} \cite{baik2005phase,rattray}:
\begin{itemize}
\item If $r<s^2$ (weak noise regime), the largest eigenvalue is well above the 'bulk' of eigenvalues due to finite sampling, and the largest eigenvector $|v_1\rangle$ has a finite overlap $\langle e| v_1\rangle$ with the principal component $|e\rangle$.
\item If  $r>s^2$ (strong noise regime), the principal eigenvalue is inside the Marcenko-Pastur 'bulk' of eigenvalues, and $|v_1\rangle$ is merely  noise, i.e. the overlap $\langle e|v_1\rangle$ vanishes in the large size limit.
\end{itemize}
In summary, recovering the principal component is impossible unless $n^\star \sim \frac{N}{s^2}$ examples at least are presented, after which the error decays monotonously; hence the name of retarded learning \cite{watkin1994optimal} coined in a slightly different context. This computation can be generalized for any finite number of eigenvalues $K>1$, associated to the set $s_1 > s_2 >... >s_K$; each time the noise level $r$ crosses $s_k^2$, one more eigenvalue pops out of the noisy bulk of eigenvalues, and the corresponding eigenvector is informative about the $k^{th}$ principal component to be inferred. A practical application of this computation is to serve as a guideline for how many principal components one should keep when PCA is used for dimensionality reduction. For instance, one can choose to keep only the eigenvalues that are larger than $\lambda_+$, the bulk top eigenvalue in eqn (\ref{mp}). 

\subsection{Incorporating prior information}
\begin{figure}
\begin{center}
\includegraphics[scale=0.5]{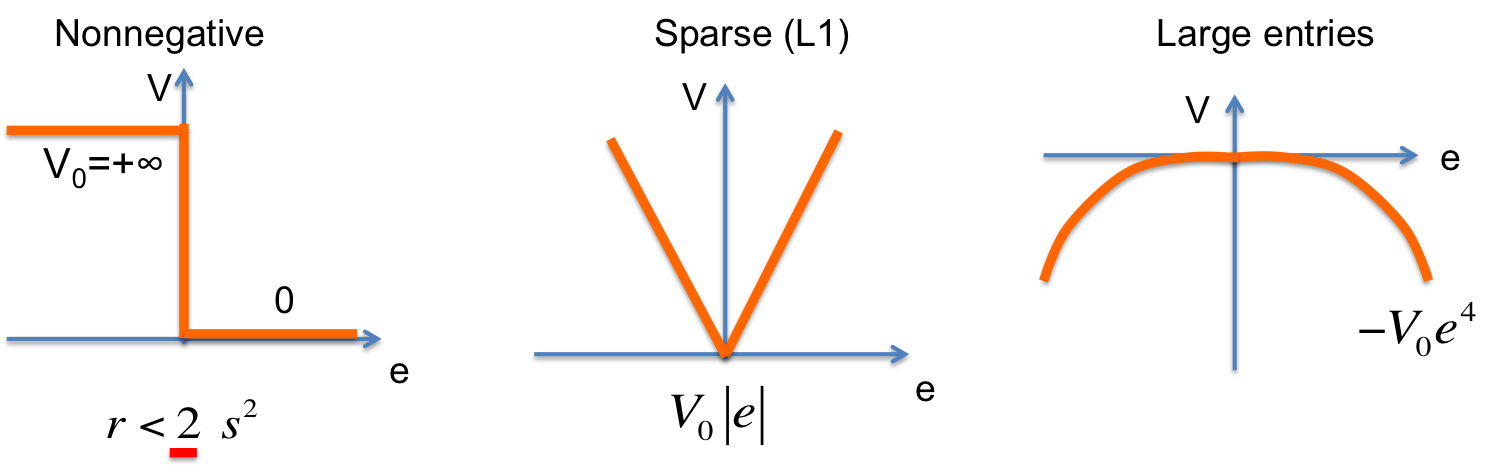}
\end{center}
\caption{Prior potentials $V(e)$ used for learning top components. In the nonnegative case (left), the top component can be inferred when $r<2s$ in the presence of the prior, which is twice bigger than the maximum-likelihood threshold, $r=s^2$. Other prior potentials include the $L_1$ regularization (middle), and a potential favoring large components (right), see \cite{monasson2015estimating}. }
\label{priorV}
\end{figure}

We have seen in the previous Section that, when $r<s^2$, it is possible to extract a vector with a finite scalar product with the top component $|e\rangle$. One natural question is whether exploiting prior information about the structure of the top component can help us increase this threshold, i.e. find out the top component with less data. We assume a prior distribution for the entries of the top component:
\begin{equation}
P(|e\rangle) =\prod_i P(e_i)\ , \quad \text{with}\quad P(e_i) \propto \exp \big[  V(e_i) \big] \ .
\end{equation}
Several expressions of interest can be considered for the potential $V$, with the representative curves shown in Fig.~\ref{priorV}. We may for instance know that the top component has all its components $e_i$ positive or zero. 
\begin{equation} \label{nonnegative}
V(e) = \left\lbrace 
\begin{array}{rr}
 +\infty \text{   if  }e < 0 \\
  0 \text{   if  }e \geq 0 \\
  \end{array}
\right.
\end{equation}
This can be useful in practice if we look for a collective excitatory mode, e.g. in gene expression data \cite{badea2005sparse,zass2007nonnegative}.
At first sight, it seems trivial to find a vector with a positive dot product with $|e\rangle$, as a good candidate is $|v\rangle = \frac 1{\sqrt N} (1,1,...,1)$. However, since $|e\rangle$ may be arbitrarily sparse (have all components equal to zero but a finite number), there is no guarantee that $\langle e|v\rangle$ is actually finite in the large $N$ limit. It was recently shown that maximizing (\ref{maxL}) under the condition $e_i\ge 0, \forall i$ leads to an estimate of the top component with a positive dot product as long as $r< 2\, s^2$ \cite{richard2014statistical}, see Fig.~\ref{priorV}. Hence, incorporating prior information about the non-negativity of the entries of the top components allows for doubling the noise level.

\begin{figure}
  \begin{minipage}[c]{0.55\textwidth}
    \includegraphics[width=\textwidth]{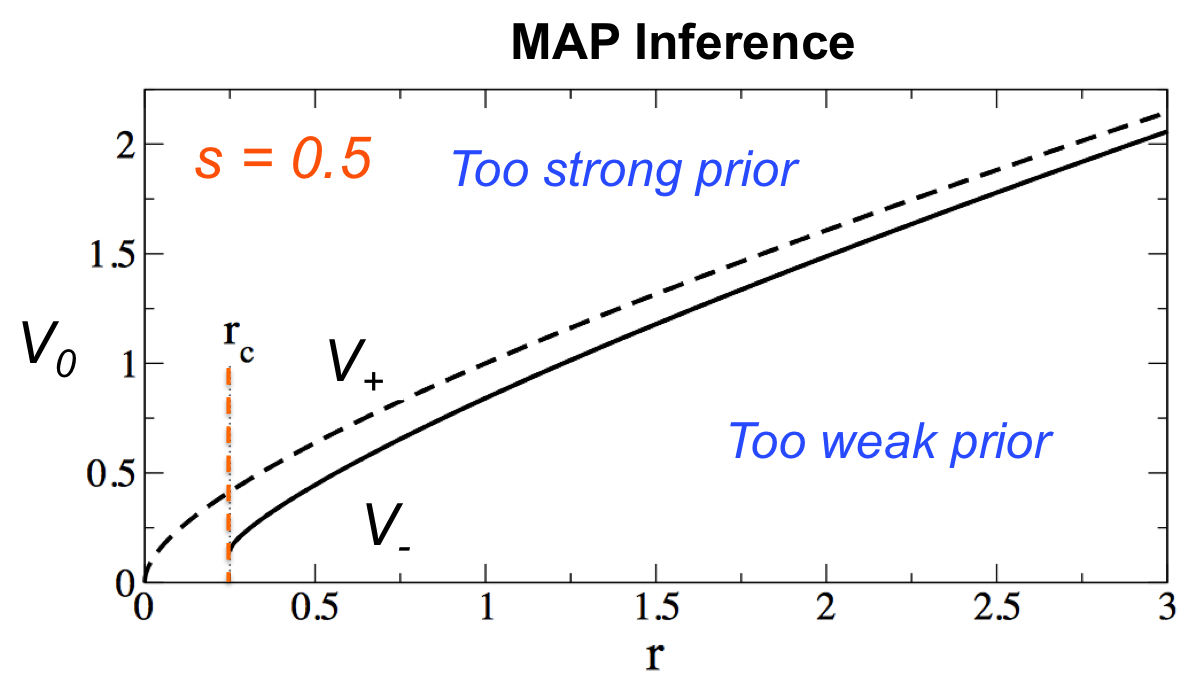}
  \end{minipage}\hfill
  \begin{minipage}[c]{0.4\textwidth}
    \caption{The PCA retarded-learning phase diagram using priors for large entries in eqn (\ref{large_entries}), for $s=0.5$. At fixed $r$, inference of the top component is possible if the strength of the large-component prior, $V_0$, is comprised between $V_-$ and $V_+$. For $V_0<V_-$, the prior is too weak, and the situation is similar to maximum likelihood decoding. For $V_0>V_+$, the prior is too strong: the inferred vector will have few large entries $e_i$, as required, but the sites $i$ carrying these large entries will not match their counterparts in the true top component.}
       \label{PCA_prior}
  \end{minipage}
\end{figure}

Other cases of interest are the $L_1$ regularization,
\begin{equation} \label{sparseV}
V(e) = V_0 \, |e|
\end{equation}
which favors sparsity. Recently, motivated by the study of covariation in protein families and the search for eigenvectors of the residue-residue correlation matrix with strong components on sites in contact on the 3D structure \cite{cocco2013principal}, we have considered the following potential \cite{monasson2015estimating,monasson2016inference}.
\begin{equation} \label{large_entries}
V(e) = - V_0\,  e^4 \ ,
\end{equation}
which favors large components. As the total normal of $|e\rangle$ is still fixed to unity, only a finite number of components can be large (and finite). Note that the cost of weak components is very small, hence this potential does not enforce any sparsity constraint, contrary to eqn (\ref{sparseV}). Instead of maximizing the likelihood, we now maximize the full a posteriori probability $P(|e\rangle | C) \propto P(C| |e\rangle) \times P(|e\rangle)$. The optimization can not be performed analytically anymore, but the analysis of the typical properties of the solution can be analyzed with the replica method \cite{mezard1987spin}. For a prior \ref{large_entries}, it is shown in particular that small values of $V_0$ can reduce the learning lag, whereas too large values impeach learning, see Fig.~\ref{PCA_prior}.

\subsection{Inverse Hopfield problem}
\label{sechopfieldinv}

PCA has also a strong connection with the inverse Ising problem of Section \ref{secising}, when the coupling matrix $J$ is constrained to have low rank, typically $\ll p$. This assumption may help avoid overfitting the data \cite{cocco2011high,cocco2013principal}. We therefore write the interaction matrix as follows,
\begin{equation}\label{pattern}
J_{ij}  = \frac 1N 
\sum_{\mu =1}^k \xi_i^\mu \, \xi_j^\mu -\frac 1N
\sum_{\mu =1}^{\hat k} \hat \xi_i^\mu \, \hat \xi_j^\mu  \ .
\end{equation}
Here, $k$ and $\hat k$ are, respectively the numbers of positive and negative eigenvalues of $J$, and the total rank  is  $k+\hat k$; note that the number of variables . The interaction matrix (\ref{pattern}), together with the Gibbs measure in ean (\ref{pisi}), define a generalized Hopfield model, made of the standard {\em attractive} patterns $\boldsymbol\xi^\mu$ and of {\em repulsive} pattern ${\hat {\boldsymbol \xi}}^\mu$,. To make the meaning of these patterns more explicit, we rewrite the probability  distribution \ref{pisi} of this generalized Hopfield model:
\begin{equation}\label{hopfieldp}
P\big({\bf s}|h,\boldsymbol\xi, \hat{\boldsymbol \xi}\big) = \frac 1{Z(h,\boldsymbol\xi, \hat{\boldsymbol \xi})} \; \exp \left({\bf h}\cdot {\bf s} +
\frac 1{2N} \sum  _{\mu=1}^k \big( {\boldsymbol\xi} ^\mu \cdot {\bf s}\big) ^2 -\frac 1{2N} \sum  _{\mu=1}^{\hat k} \big({\hat {\boldsymbol\xi}} ^\mu \cdot {\bf s}\big)^2 \right) \ ,
\end{equation}
where $\cdot$ denotes the scalar product (summation over components $i$). The meaning of the patterns is transparent: they define  favored, for attractive patterns, or disfavored, for repulsive patterns,  
directions in the space of configurations $\bf s$, along which the probability increases or decreases quadratically.

Attractive and repulsive patterns may be inferred through minimization of the cross-entropy defined in eqn (\ref{entropy}). To the lowest order in $\xi_i^\mu/\sqrt N$ and $\hat \xi_i^\mu/\sqrt N$, one finds that the fields and patterns minimizing $S$ are given by
\begin{eqnarray}\label{Ninfini}
h_i&=& \log p_i \nonumber \\
\xi _i ^\mu &=& \sqrt{1 - \frac 1{\lambda ^\mu}}\ 
\frac{v_i^\mu} {\sqrt{p_i (1-p_i)}} \qquad (\mu=1,\ldots , k)\nonumber \\
\hat \xi _i ^\mu &=& \sqrt{\frac 1{\lambda ^{N-\mu}}-1}\ 
\frac{v_i^{N -\mu}} {\sqrt{p_i (1-p_i)}}\qquad (\mu=1,\ldots , \hat k)
\end{eqnarray}
where $\lambda ^1\ge \lambda^2\ge  ... \ge 1\ge ...\ge \lambda ^{N-1}\ge\lambda ^{N}$ are the eigenvalues of the Pearson correlation matrix,
\begin{equation}\label{matrixc}
C_{ij}= \frac{p_{ij} -p_i\,p_j}{\sqrt{p_i(1-p_i)\, p_j(1-p_j)}} \ ,
\end{equation}
and the ${\bf v}^\mu$ are the associated eigenvectors (with squared norms equal to $N$)\footnote{Note that the Hopfield model is, by construction, invariant under global rotations in the pattern index space, e.g.  any rotation ${\cal O}$ of all the attractive patterns in the $k-$dimensional space:
\begin{equation}
\boldsymbol \xi ^\mu \to \sum _{\nu}{\cal O}^{\mu, \nu} \; \boldsymbol \xi ^\nu \ .
\end{equation}
In other terms, the patterns are defined up to a rotation and are not unique; the gauge chosen in eqn (\ref{Ninfini}) corresponds to orthogonal patterns in site space. Obviously,  the couplings $J_{ij}$ are gauge-invariant.}.

The above procedure is strongly reminiscent of PCA. Formula (\ref{Ninfini}) shows, however, that the patterns do not coincide with the eigenvectors of $C$ due to the presence of $p_i$-dependent terms. Furthermore, the presence of the $\lambda^\mu$-dependent factor discounts the patterns corresponding to eigenvalues close to unity. This effect is easy to understand in the case of independent spins: in the limit of perfect sampling ($B\to\infty$), $C$ coincides with the identity matrix, hence $\lambda^\mu=1, \forall \mu$, and the patterns and the couplings vanish as they should. In the general case of coupled spins, the sum of the eigenvalues of $C$ is equal to $N$ (since $C_{ii}=1, \forall i$). Therefore, the largest and smallest eigenvalues are guaranteed to be, respectively, above and below unity, and the corresponding attractive and repulsive patterns are real valued.

\begin{figure}
  \begin{minipage}[c]{0.58\textwidth}
    \includegraphics[width=\textwidth]{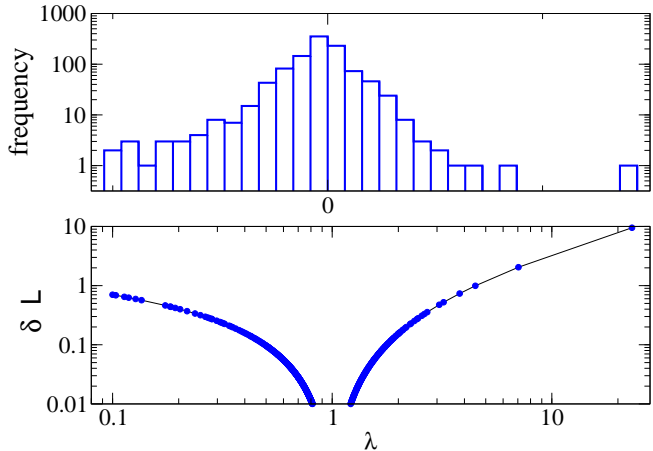}
  \end{minipage}\hfill
  \begin{minipage}[c]{0.4\textwidth}
    \caption{Top: Eigenvalue spectrum of the Pearson correlation matrix for  the 
sequences of the trypsin inhibitor family (PF00014); the noise ratio is $r\simeq 0.5$, hence the edges of the Marcenko-Pastur spectrum are approximately $\lambda_-=0.09$ and $\lambda_+=2.9$, see eqn (\ref{mp}). Bottom:  pattern
contributions to the log-likelihood of the inferred Hopfield model for the
patterns corresponding to the eigenvalues along the $x$-axis.
The most-contributing patterns  are attractive patterns corresponding to
the largest eigenvalues and repulsive patterns corresponding to the
smallest eigenvalues. See \cite{cocco2013principal} for more details. }
       \label{PCA14}
  \end{minipage}
  \vskip 0cm
\end{figure}

  \begin{figure}
\begin{center}
{\includegraphics[width=12cm]{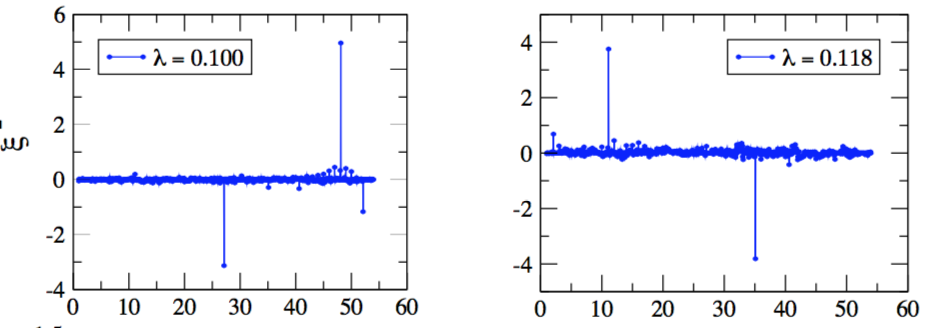}}
\caption{ Two of the smallest-eigenvalue repulsive patterns obtained for the trypsin inhibitor family (PF00014). $x$ index: $i+s/20$, where $i$ is the site index and $s=1\ldots 20$ is the amino acid (or gap) index.  Each pattern is localized on essentially two components, corresponding to two sites $i_1,i_2$ in contact through a cysteine-cysteine bridge. Importantly, the two sites are close on the 3D fold but distabt along the protein backbone. See \cite{cocco2013principal} for more details.}
\label{fig-schema-2}
\end{center}
\end{figure}

Inserting expression (\ref{Ninfini}) in the cross-entropy (\ref{entropy}), we obtain the contribution (per data configuration) of pattern $\mu$ to the log-likelihood,
\begin{equation}\label{gainphi}
\delta {\cal L} ^\mu = \frac 12 \big( \lambda^\mu -1 -\log \lambda^\mu \big) \ ,
\end{equation}
a quantity which is strictly positive for $\lambda^\mu \ne 1$, see Fig.~\ref{fig-schema-2}. This expression helps select most relevant patterns, in decreasing order of their contributions $\delta {\cal L}^\mu$. This is in analogy with PCA when one selects the signal eigenvectors as the ones detaching most from the spectrum of the  Marcenko-Pastur distribution, see Section \ref{secretard}. However, at difference with PCA, where only top eigenvalues (large $\lambda^\mu$) and attractive patterns are taken into account, selected patterns in the inverse Hopfield  model are on both ends of the spectrum. Small eigenvalues, much bloe $\lambda=1$, can give large contributions to the log-likelihood (Fig.~\ref{fig-schema-2}). In applications to the study of covariations in protein families, repulsive patterns can be shown to be localized on a small number of sites; they are much more information about structural constraints in the protein, e.g. on the pairs of amino acids in contact \cite{cocco2013principal}.

\section{Compositional Representations: Restricted Boltzmann Machines (RBM)}
\subsection{Definition and motivation}
\begin{figure}[b]
  \begin{minipage}[c]{0.3\textwidth}
    \includegraphics[width=\textwidth]{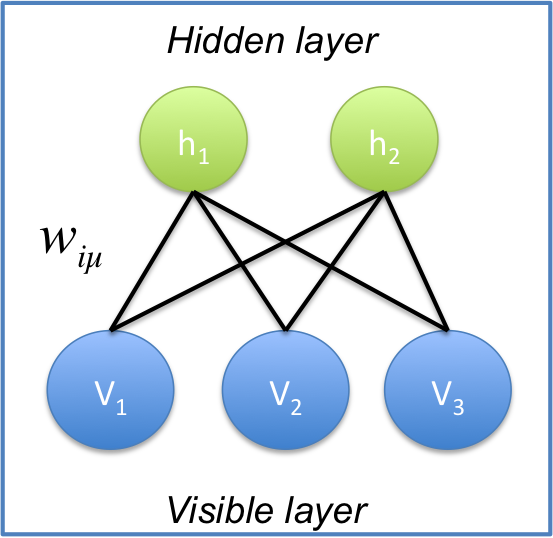}
  \end{minipage}\hfill
  \begin{minipage}[c]{0.55\textwidth}
    \caption{Architecture of Restricted Boltzmann Machines. A RBM is defined on a bidirectional bipartite graph, with a visible (v) layer that represents the data, connected to a hidden (h) layer supposed to extract statistically meaningful features from the data and, in turn, to condition their distribution. There are $N$ visible units indexed by $i$, and $M$ hidden units indexed by $\mu$. The connections between visible and hidden units are denoted by $w_{i\mu}$.}
       \label{fig:archi}
  \end{minipage}
\end{figure}

A Restricted Boltzmann Machine (RBM) is a graphical model, i.e., a probability distribution over a multidimensional data set, similar to the multivariate gaussian distribution or the Boltzmann Machine distribution. It is constituted by two sets of random variables, a visible layer (v) -the data layer- and a hidden layer (h), which are coupled together, see Fig.~\ref{fig:archi}. The joint probability distribution of the visible and hidden unit configurations, ${\bf v} = (v_1,v_2,...,v_N)$ and  ${\bf h} = (h_1,h_2,...,h_M)$, is the Gibbs distribution
\begin{equation}
P(\boldsymbol v,\boldsymbol h) = \frac{1}{Z} e^{-E(\boldsymbol v,\boldsymbol h)} \ ,
\end{equation}
defined by the energy
\begin{equation}
E(\boldsymbol v,\boldsymbol h) = \sum_{i=1}^N \mathcal{U}_i(v_i) + \sum_{\mu =1}^M \mathcal{U}_\mu(h_\mu)  - \sum_{i,\mu} w_{i,\mu} v_i h_\mu \ ,
\end{equation}
where the  $\mathcal{U}_i, \mathcal{U}_\mu$ are unary potentials that control the marginal distributions of the variables $v_i, h_\mu$, and the weight matrix $w_{i,\mu}$ couples the visible and hidden layers. Depending on the choice of the potentials, the visible and hidden variables can be binary or continuous. The visible potentials $\mathcal{U}_i$ is in general chosen based on the data we want to model; for example if $v \in [0,1]$, then $U_i(v_i) = -g_i v_i$, where the field $g_i$ is a parameter of the model. The hidden potentials $\mathcal{U}_\mu$ can be chosen arbitrarily as long as sampling is feasible. Some useful examples are:
\begin{itemize}
\item The Bernoulli potential: $\mathcal{U}_\mu(h_\mu) = -g_\mu h_\mu \quad \text{with} \quad h_\mu \in [0,1]$\ ;
\item The Quadratic potential: $\mathcal{U}_\mu(h_\mu) = \frac 12 \, h_\mu^2, \; \; h_\mu \in \mathbb{R}$ \ ;
\item The ReLU potential:
$\mathcal{U}_\mu(h_\mu) = \left\lbrace  \begin{array}{r r r} 
\frac 12\, { h_\mu^2}+ \theta_\mu \, h_\mu &  \text{if  } & h_\mu \geq 0 \\
+\infty & \text{if  } & h_\mu < 0
\end{array} \right.$ \ .
\end{itemize}

By marginalizing over the hidden units, one can compute the probability distribution over the visible layer:
\begin{equation} \label{marginal}
P(\boldsymbol v) = \int \prod_\mu \, dh_\mu\; P(\boldsymbol v,\boldsymbol h) \equiv \frac{1}{Z_{eff}} \exp \left[ - E_{eff}(\boldsymbol v) \right]
\end{equation}
The former expression \ref{marginal} can be expressed analytically  in terms of the weight matrix and the potentials. Training an RBM consists in fitting this marginal distribution to the data by maximum likelihood \cite{smolensky1986information}. Unlike multivariate Gaussian distributions studied in the previous Section, the optimal RBM must be found numerically, e.g. using approximate stochastic gradient ascent over the likelihood \cite{fischer2014training}.

\begin{figure}
\includegraphics[width=.42\textwidth]{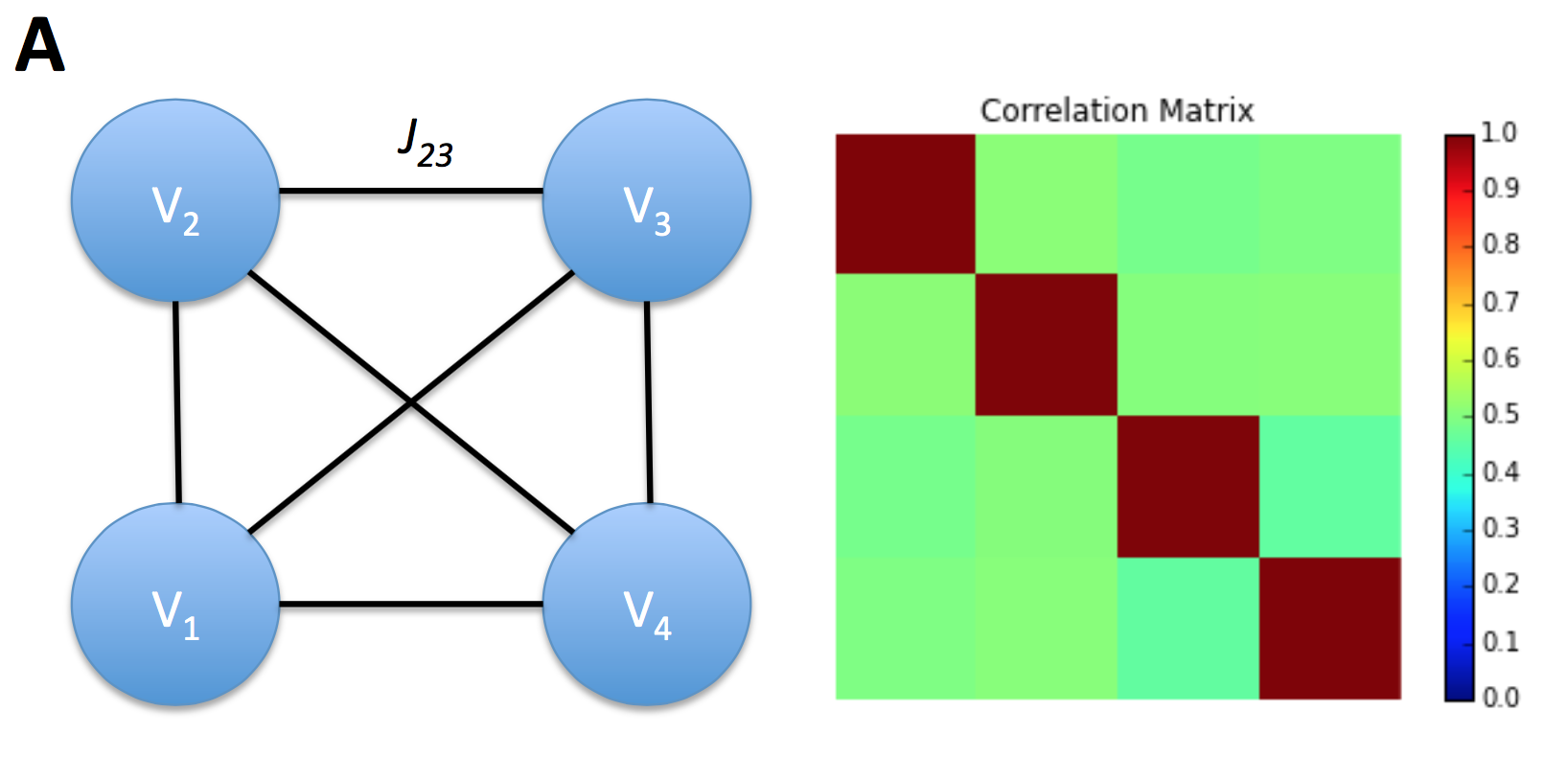} \hskip.5cm
\includegraphics[width=.55\textwidth]{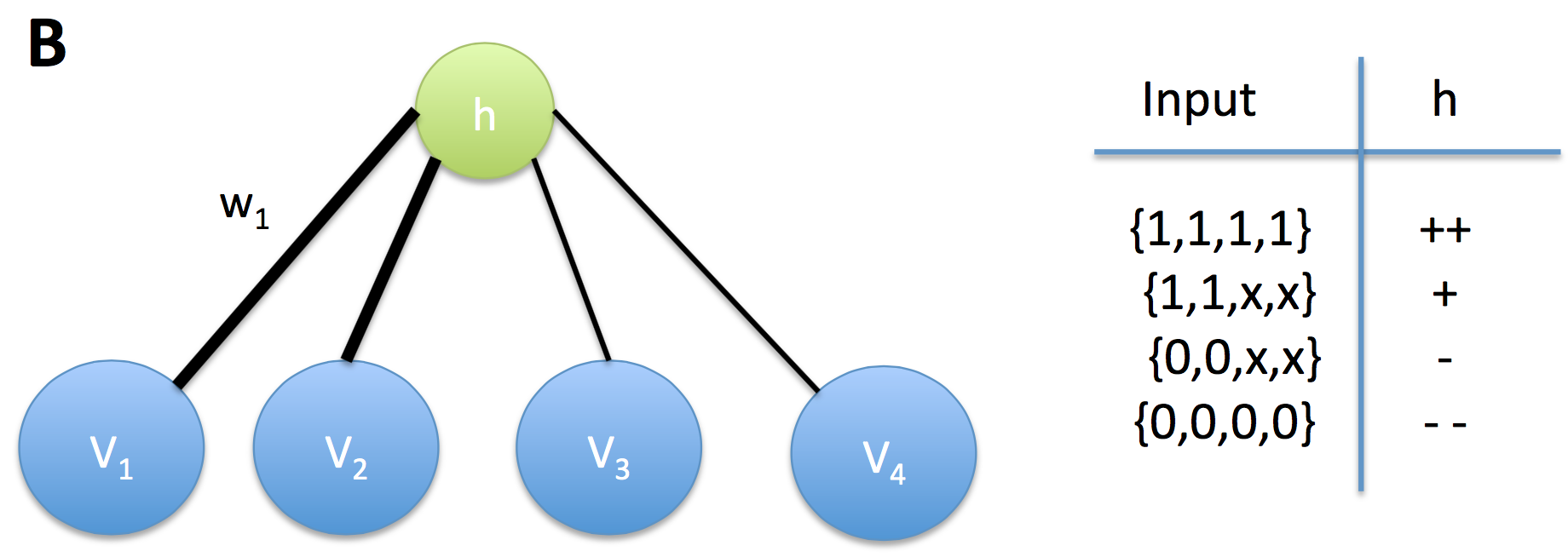}
\caption{How to model correlations among a set of variables. {\bf A.} Boltzmann Machine approach: The matrix of pairwise correlations between variables is computed from data, and a network of couplings is inferred to reproduce those correlations. {\bf B.} Restricted Boltzmann Machine approach: observed correlations are due to one or more common input(s), whose values drive the configurations of the variables. A network of connection between the visible layer (support of data configurations) and a layer of hidden units (support of common inputs) is found to maximize the probability of the data items. The rightmost column indicates the magnitude $h$ of the hidden unit as a function of the visible configuration.}
 \label{RBM_intuition}
\end{figure}

We stress that, in contradistinction with  Boltzmann Machines (Ising models) or multivariate Gaussian distributions, there are no direct couplings between pairs of units in the same layer (hence, the name restricted). RBM can nonetheless model correlations between visible variables, as the latter can be indirectly correlated through the hidden layer. Informally speaking, instead of explaining the correlations between several visible units through a set of adequate couplings, we interpretate them as collective variation driven by the common inputs (the hidden units) shared by these visible units, see Fig.~\ref{RBM_intuition}. The hidden units thus represent collective modes of variation of the data. 

Another way to state this is to observe that, as one marginalizes over the hidden layer, effective couplings between visible layer units arise. For instance, it is easy to show that for Gaussian hidden units, i.e. for the Quadratic potential ${\cal U}_\mu$ in the list above, the marginal distribution over the visible layer is:
\begin{equation}
E_{eff}(\boldsymbol v) = -\sum_i g_i \,v_i + \frac{1}{2} \sum_\mu \left( \sum_i w_{i\mu} v_i \right)^2
\end{equation}
In that case, we recognize a pairwise effective Hamiltonian, the Hopfield model with $M$ patterns \cite{Hopfield82,barra12}. In general, non-quadratic hidden-unit potentials generate effective Hamiltonian for the visible units with high-order interactions. The presence of couplings to all orders produced from a unique set of $N\times M$ connections $w_{i\mu}$ has deep effects on the sampling dynamics of RBM.

\subsection{Sampling}
\label{secsam}
\begin{figure}
  \begin{minipage}[c]{0.7\textwidth}
    \includegraphics[width=\textwidth]{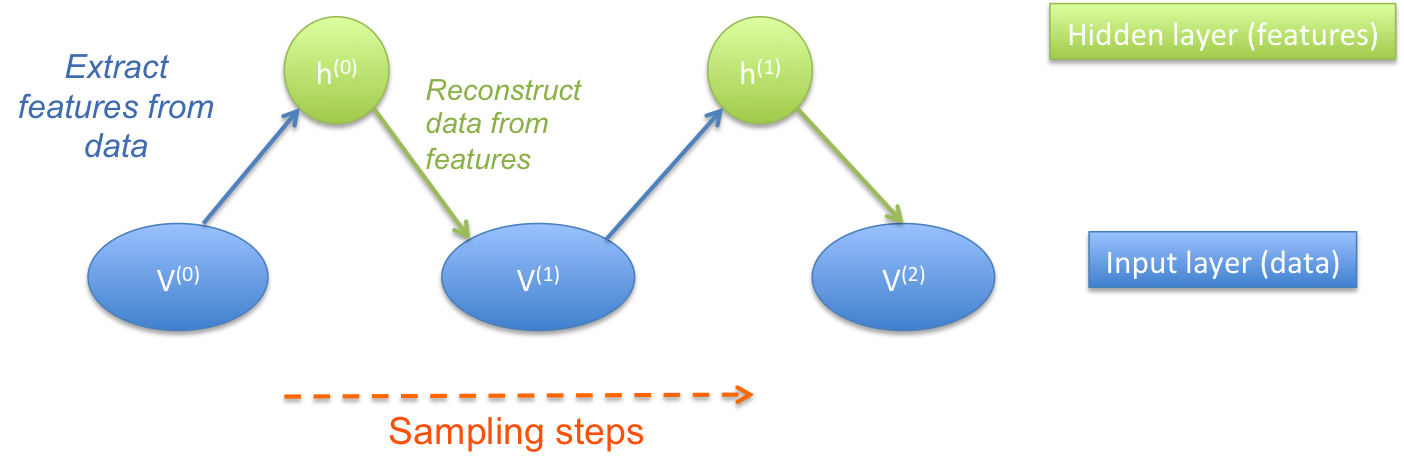}
  \end{minipage}\hfill
  \begin{minipage}[c]{0.25\textwidth}
    \caption{Back-and-forth sampling procedure in RBM. Hidden configurations $\bf h$ are sampled from visible configurations $\bf v$, and, in turn, define the distribution of visible configurations at the next sampling step.}
       \label{sampling_RBM}
  \end{minipage}
\end{figure}

\begin{figure}
  \begin{minipage}[c]{0.4\textwidth}
    \includegraphics[width=\textwidth]{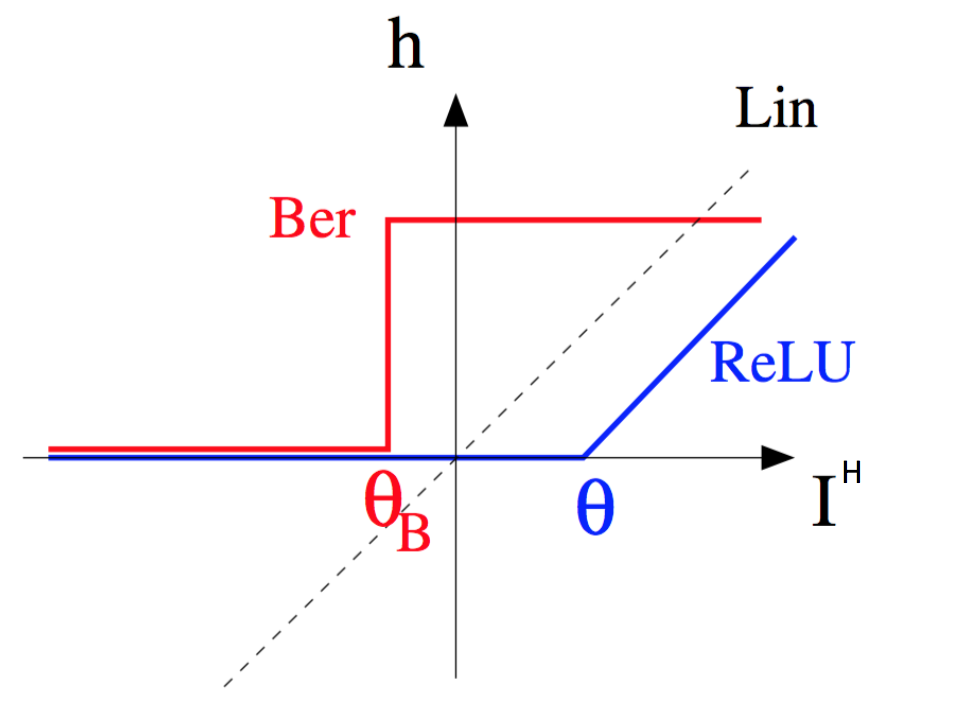}
  \end{minipage}\hfill
  \begin{minipage}[c]{0.5\textwidth}
    \caption{Transfer functions $\Phi$ for various hidden units potentials ${\cal U}_\mu$, see main text. The transfer function gives the most likely value of the hidden unit, $h^*$, as a function of the input $I^H$ received from the visible units.}
\label{transfer_functions_RBM}
  \end{minipage}
\end{figure}

The connection with data representation algorithms is best seen when considering the sampling scheme. Since there are no connections within a layer, the hidden layer units are conditionally independent given the configuration of the visible layer, and conversely; hence the following Gibbs sampling procedure, schematized in Fig.~\ref{sampling_RBM}:
\begin{itemize}
\item Compute hidden units inputs $I_\mu^H = \sum_i w_{i\mu} v_i$
\item Sample each hidden unit independently $P(h_\mu | I_\mu^H) \propto \exp \left[ - U_\mu(h_\mu) + h_\mu I_\mu^H \right]$
\item Compute the visible layer inputs $I_i^V = \sum_\mu w_{i\mu} h_\mu$
\item Sample each visible unit independently $P(v_i | I_i^v) \propto \exp \left[ (g_i + I_i^V) v_i \right]$
\end{itemize}
The first two steps can be seen as a stochastic feature extraction from configuration ${\bf v}$, whereas the last two steps are a stochastic reconstruction of ${\bf v}$ from the features ${\bf h}$. One can define in particular a data representation as the most likely hidden layer configuration given a visible layer configuration, that is, through the set of 
\begin{equation}
{h^*_\mu}({\bf v}) = \arg \max P(h_\mu |{\bf v}) = \argmaxB P(h_\mu|{\bf v})= \Phi_\mu(I_\mu^H({\bf v})) \ ,
\end{equation}
where $\Phi_\mu = (\mathcal{U}_\mu')^{-1}$ is the transfer function, see Fig.~\ref{transfer_functions_RBM}.

\subsection{Phenomenology of RBM trained on data} 
\label{phenomenology}
\begin{figure}
  \begin{minipage}[c]{0.6\textwidth}
    \includegraphics[scale = 0.52]{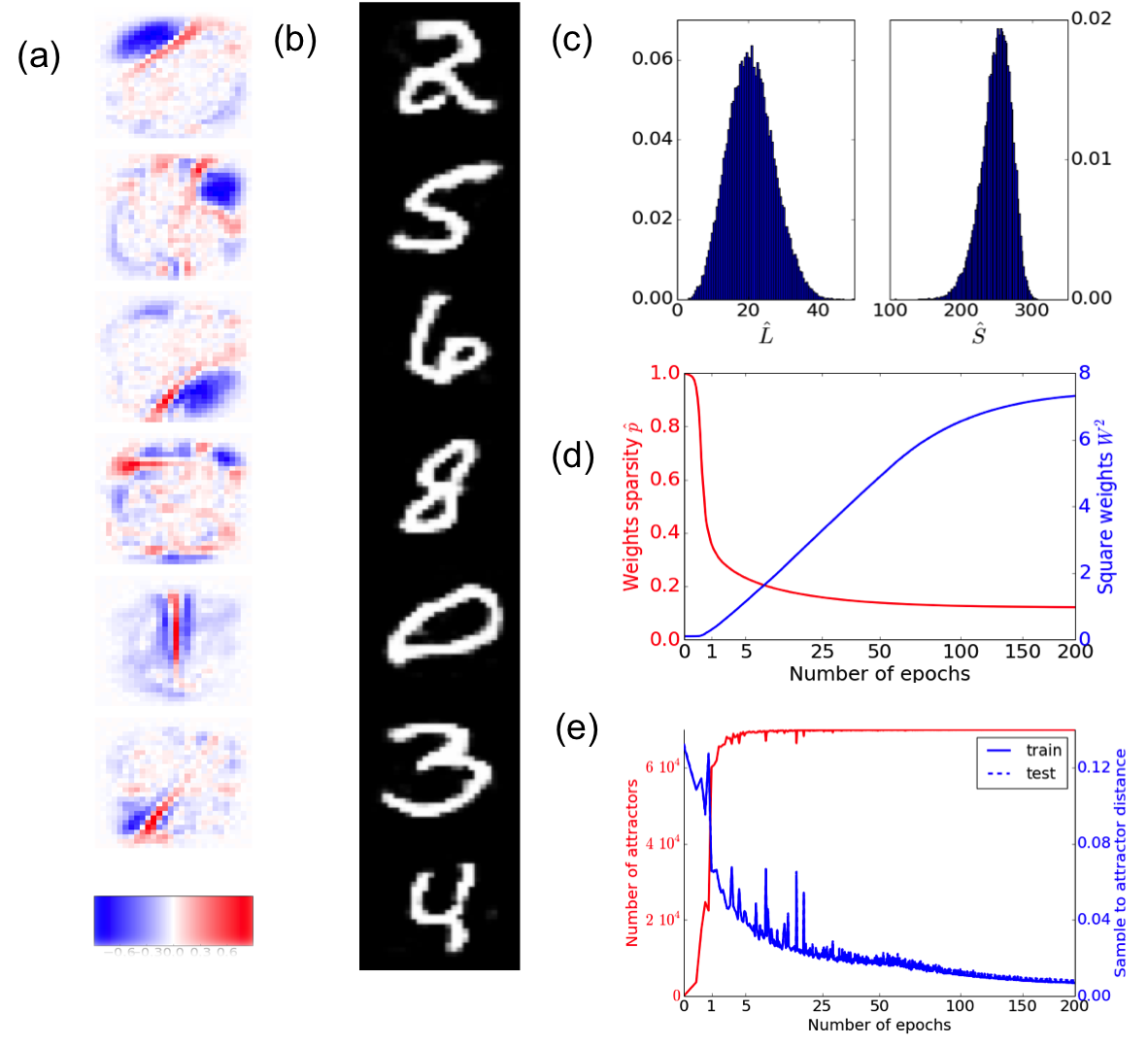}
  \end{minipage}\hfill
  \begin{minipage}[c]{0.35\textwidth}
\caption{Training of RBM on MNIST. Data are composed of $N=28\times 28$ binarized images, and the RBM includes $M=400$ hidden ReLU.  {\bf (a)} Set of weights ${\bf w}_{\mu}$ attached to  four  representative hidden units $\mu$. {\bf (b)}  Averages of $\bf v$ conditioned to five hidden-unit configurations $\bf h$ sampled from the RBM at equilibrium. Black and white pixels correspond respectively to averages equal to $0$ and 1; few intermediary values, indicated by grey levels, can be seen on the edges of digits.  {\bf (c)}  Distributions of the numbers of very strongly activated hidden units, $\hat L $ (left), and  of silent hidden units, $\hat S$ (right), at equilibrium. {\bf (d)} Evolution of the weight sparsity $\hat p$  (red) and the squared weight value $W_2$ (blue). The training time is measured in epochs (number of passes over the data set), and represented on a square--root scale. {\bf (e)} Evolution of the number of distinct local maxima of $P({\bf v})$ in eqn (\ref{marginal}) (left scale) and distance to the original sample (right scale, for training and test sets). For each sample, the local minimum is obtained through $T=0$--sampling  of the RBM, see Section \ref{secsam}.}
\label{phenomenology_RBMs}
  \end{minipage}
\end{figure}

Once maximum-likelihood training is completed, RBM can be very good generative models for complex, multimodal distributions. In the following, we describe the phenomenology of RBM, with various kind of hidden-unit potentials, trained on MNIST, a dataset of 60,000 $28\times28$ grayscale images of handwritten digits. Each image can be flattened, binarized by thresholding the grayscale level, into a 784-dimensional binary vector. The following observations can be done:
\begin{itemize}
\item After training, samples drawn from the equilibrium distribution of Bernoulli or ReLU RBM look like real digits, suggesting that it is a good fit for the data distribution, see Fig.~\ref{phenomenology_RBMs}(b). On the contrary, Gaussian RBM, \textit{i.e.} pairwise Hamiltonians, do not fit the data distribution as well.
\item Each hidden unit is activated selectively by the presence of a specific feature of the data: this is is seen by visualizing the columns of the weight matrix $w_{i\mu}$, see Fig.~\ref{phenomenology_RBMs}(a). The features are strokes, that is, small part of digits. The weight matrix is therefore essentially sparse, with a fraction of nonzero weights $p \sim 0.1$, see Fig.~\ref{phenomenology_RBMs}(d).
\item For ReLU hidden units, each data image strongly activates around $\sim 20$ hidden units, whereas most hidden units are silent or weakly activated, see Fig.~\ref{phenomenology_RBMs}(c). For a precise definition of how the number strongly hidden units is estimated, see eqn (\ref{pr}) and \cite{tubiana2017emergence}.
\item The learnt probability distribution is very rough, with many local maxima of probability (much larger than the values of $N$ or $M$), as seen in Fig.~\ref{phenomenology_RBMs}(e). Remarkably, after training, each data sample is within few pixels of a local maximum of probability. This shows the combinatorial nature of RBM, capable of generating a very large number of configurations after training.
\end{itemize}
This phenomenology raises several questions. First, how can such simple networks generate a complex distribution with a large variety of local minima, matching the original data points? Secondly, why do some hidden unit potentials give good results, whereas others do not? Lastly, can we connect this behavior to the one of the Hopfield model, corresponding to the case of quadratic potential?

\subsection{Statistical mechanics of RBM}
\begin{figure}
  \begin{minipage}[c]{0.6\textwidth}
    \includegraphics[width=\textwidth]{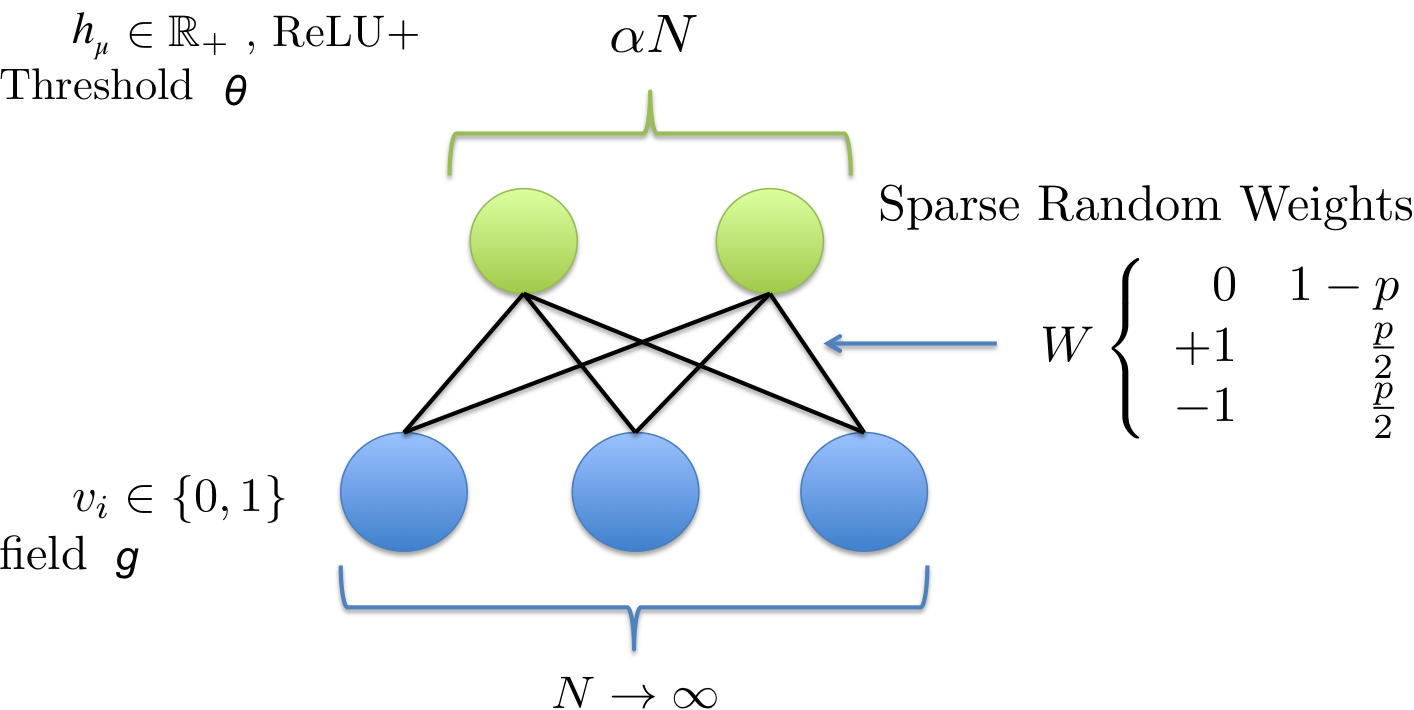}
  \end{minipage}\hfill
  \begin{minipage}[c]{0.3\textwidth}
    \caption{The Random-RBM ensemble, with its control parameters: threshold $\theta$ of hidden ReLU, ratio $\alpha$ of the sizes of the hidden and visible layers, field $g$ on visible units, sparsity $p$ of the weights (rescaled by $W=1/\sqrt N$). }
\label{RBM_drawing}
  \end{minipage}
\end{figure}

It is hopeless to provide answers to these questions in full generality for a given RBM with parameters fitted from real data. However, statistical physics methods and concepts allow us to study the  typical energy landscape and properties of RBM drawn from appropriate random ensembles. We follow this approach hereafter, using the replica method \cite{tubiana2017emergence}. We define the Random-RBM ensemble model for ReLU hidden units as follows, see also drawing in Fig.~\ref{RBM_drawing},
\begin{itemize}
\item $N$ binary visible units, $M$ ReLU hidden units, with $N,M \rightarrow \infty$ and $\alpha = \frac{M}{N}$ is finite.
\item uniform visible layer fields, i.e. $g_i = g,\ \forall i$.
\item uniform hidden layer thresholds, i.e. $\theta_\mu = \theta, \ \forall \mu$.
\item a random weight matrix $w_{i\mu} = \frac{\xi_{i\mu}}{\sqrt{N}} $, where each 'pattern' $\xi_{i\mu}$ is drawn independently, taking values $+1,-1$ with probabilities $\frac{p}{2}$ and $0$ with probability $1-p$. The \textit{degree of sparsity} $p$ is the fraction of non-zero weights. 
\end{itemize}
Hence, $\alpha$, $p$, $g$ and $\theta$ are the control parameters of our model. Several variants or special cases have already been addressed in the literature. Choosing Gaussian hidden units and $\pm 1$ visible units leads back to the original Hopfield model, studied in \cite{AGS87}. The sparse weight distribution was previously introduced to study parallel storage of multiple sparse items in the Hopfield model \cite{agliari2012multitasking,agliari2013immune}. 

It is important to understand the magnitude of hidden-unit activations for a given a visible layer configuration $\bf v$. Two cases are encountered:
\begin{itemize}
\item let us call $L$ the number of hidden units $\mu$ coding for features ${\bf w}_\mu$ present in $\bf v$. These hidden units will be strongly activated, as their inputs $I^H_\mu= {\bf w}_\mu\cdot {\bf v}$ will be strong and positive, comparable to the product of the norms of ${\bf w}_{\mu}$, of the order of $\simeq \sqrt p$ for large $N$, and $\bf v$, of the order of $\sqrt {p\,N}$. Therefore, we expect  $I^H_\mu$ to scale as $m\sqrt N$, where the prefactor $m$, called magnetization, is finite in the thermodynamical limit. 
\item The remaining $M-L$ hidden units $\mu'$ have, however, features ${\bf w}_{\mu'}$ essentially orthogonal to $\bf v$. Hence, the vast majority of hidden units receive random inputs $I^H_{\mu'}$ fluctuating around zero, with finite variances. 
\end{itemize}

\begin{figure}
\begin{center}
\includegraphics[scale=0.35]{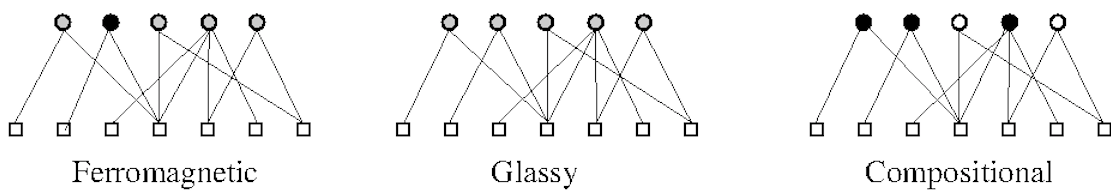}
\end{center}
\caption{The three regimes of operation of Random RBM, see text. Black, grey and white hidden units symbolize, respectively, strong ($h\sim \sqrt N$), weak ($h\sim \pm 1$) and null ($h=0$) activations.}
\label{all_phases}
\end{figure}

To answer the questions raised in Section \ref{phenomenology}, we are interested in computing the averages of $m$, $L$ over the distribution and over the random weights. They can be obtained through a  replica computation of the average free energy,
\begin{equation}
f(\alpha,p,g,\theta)  \equiv \lim_{N \rightarrow \infty} -  \frac{1}{\beta N}\; \overline{ \log Z \left(\alpha,\beta,p,g,\theta, \{\xi_{i\mu} \} \right) } \ ,
\end{equation}
where the overbard denotes the average over the $ \{\xi_{i\mu} \} $ and the partition function reads
\begin{equation}
Z \left(\alpha,\beta, p,g,\theta,\{\xi_{i\mu}\}\right) = \sum_{{\bf v} \in \{0,1\}^N} \int \prod_{\mu=1}^{M} dh_\mu\; e^{-\beta E( {\bf v},{\bf h})}  \ .
\end{equation}
After some algebra, see \cite{tubiana2017emergence}, we find that $f(\alpha,p,g,\theta)$ is obtained through optimizing a free-energy functionial over the order parameters $L,m,q,r,B,C$:
\begin{itemize}
\item  $m$ and $L$ are, respectively, the magnetization and the number of feature-encoding hidden units, 
\item $r$ is the mean squared activity of the other hidden units, 
\item $q= \frac{1}{N} \sum_i \overline{\langle v_i\rangle}$ is the mean activity of the visible layer in the GS,
\item $B,C$ are response functions, {\em i.e.} derivatives of the mean  activity of, respectively, hidden and visible units with respect to their inputs. 
\end{itemize}
For non-sparse weights ($p=1$) and depending on the values of the other control parameters, the system can show one of two following qualitative different behaviors, as is found for the Hopfield model \cite{AGS87} and Fig.~\ref{all_phases}:
\begin{itemize}
\item A \underline{ferromagnetic} phase, in which hidden configurations with $L=1$ and $m>0$ dominate.  Visible configurations have strong overlap with one feature, say, $\mu=1$. It is likely that $v_i=1$ if  $\xi_{i,1}=1$ and $v_i=0$ if $\xi_{i,1}=-1$. As the choice of $\mu$ is arbitrary, there are $\alpha N$ such 'basins' of visible configurations. Phases with $L>1$, i.e. having strong overlap with several features exist and may be thermodynamically stable, but are unfavorable: their free energies increase with $L$.
\item A \underline{spin-glass} phase, in which configurations with $m=0$ dominate. Most configurations have weak overlap $\sim \frac{1}{\sqrt{N}}$ with all hidden units. 
\end{itemize}
The phase transition occurs out of the frustration in the system. Assume for instance that the system is in the ferromagnetic phase. The input recevied by a visible unit, say, $i$ has a strong contribution (of the order of $1$ as $N\to\infty$) from the strongly magnetized unit, say, $\mu=1$, and a lot (of the order of $\alpha N$) of weak inputs (of the order of $\pm 1/\sqrt N$) from the other hidden units.  As the ratio $\alpha$ increases, these numerous, weak noisy contributions win over the unique, strong signal contribution, and the systems enters the glassy phase. The transition takes place at a well defined value of $\alpha$, which depends on $\theta$ and $g$ \cite{tubiana2017emergence}.

For small $p$, a new intermediate qualitative behavior emerges:
\begin{itemize}
\item The \underline{compositional} phase, in which visible configurations have strong overlap with $L$ features, where $1\ll L\ll M$, see Fig.~\ref{all_phases}. As observed for RBM trained on real data in Fig.~\ref{phenomenology_RBMs}(e), random RBM may generate a combinatorial diversity of low-energy visible configurations,  corresponding to different choices of the subset $\{ \mu_1,... ,\mu_L\}$ of strongly activated hidden units. This new phase is found in the low $p$ limit, and for appropriate values of  the threshold $\theta$ (large enough to silence a large number of hidden units and suppress interference, see  Fig.~\ref{phenomenology_RBMs}(c)), and of the field $g$ (to reproduce the average activity of the data in the visible layer).
\end{itemize}

\subsection{Validation on data}
One of the outcomes of our statistical physics analysis is that, in the compositional phase, the number $L$ of strongly activated hidden units scales as the inverse of the degree of sparsity, $p$. More precisely, $L\sim \frac{\ell}p$, when $p\to 0$, where $\ell$ is determined by minimzing the free energy of the Random-RBM model. The minimum $\ell^*$ of the free energy is found at $\ell^*>0$ in the compositional phase, contrary  to the ferromagnetic phase, where $\ell^*=0$.

\begin{figure}
  \begin{minipage}[c]{0.5\textwidth}
    \includegraphics[width=\textwidth]{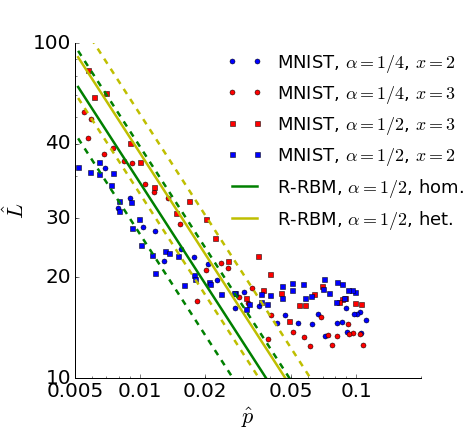}
  \end{minipage}\hfill
  \begin{minipage}[c]{0.4\textwidth}
    \caption{ Average number $L$ of active hidden units  vs.  degree $p$ of sparsity of the weights, for RBM trained on MNIST data. Values of the ratio $\alpha$ and of the exponent $x$ in the regularization term in eqn (\ref{regu}) are reported in the figure. The figure also shows the theoretical curve obtained for Random RBM ($CV=0$, as in Fig.~\ref{RBM_drawing}), and for RBM under another more realistic statistical ensemble of random weights, in which the degree of sparsity $p$ fluctuates with the visible sites $i$. Dashed lines show one standard deviations away from the mean value of $L$ due to finite-size fluctuations,  see \cite{tubiana2017emergence} for details.
    }
\label{scalingL}
  \end{minipage}
\end{figure}

This prediction can be tested in RBM trained on real data, e.g. MNIST, see Fig.~\ref{phenomenology_RBMs}(b). The value of $p$ at the end of training without any regularization was found to be $\sim 0.1$, see Fig.~\ref{phenomenology_RBMs}(d). However, higher sparsities, i.e. lower values of $p$, can be imposed through regularization of the weights. To do so, we add to the log-likelihood the penalty term
\begin{equation}\label{regu}
C(\{w_{i\mu}\}) = - \sum _\mu \big(\sum_i |w_{i\mu}|\big)^x\ ,
\end{equation}
where $x\ge 0$. The case $x=1$ gives standard $L_1$ regularization, while, for $x > 1$, the effective penalty strength, $\propto \big(\sum_i |w_{i\mu}|\big)^{x-1}$, increases with the weights, hence promoting homogeneity among hidden units. After training we generate Monte Carlo samples of each RBM at equilibrium, and monitor the average number of active hidden units, $L$, estimated through the participation ratio
\begin{equation}\label{pr}
L = \frac{(\sum_\mu h_\mu^2)^2} { \sum_\mu h_\mu^4}\ .
\end{equation}
By changing the value of $x$, we obtain, at the end of training, RBM with higher sparsities. Figure~\ref{scalingL} shows that the theoretical scaling law $L\sim \ell^*/p$ is well reproduced over one decade of variation of $p$. In addition, the product $L\times p$ is in good agreement with the theoretical prediction $\ell^*$ \cite{tubiana2017emergence}.

\vskip .3cm
\noindent {\bf Acknowledgements.} This work benefited from the financial support of the Human Frontier Science Program through the RGP0057/2016 project.

%%%%%biblio%%%%%%%

\bibliographystyle{ieeetr}  
%\bibliography{articles}
\bibliography{biblio}

\end{document}